\documentclass[english,twocolumn,floatfix,aps,showpacs,superscriptaddress,bibliography]{revtex4-1}
\usepackage[english]{babel}                                                           
\usepackage[utf8]{inputenc}                                                            
\usepackage[T1]{fontenc}
\usepackage{epsfig, mathtools}
\usepackage{amsmath}
\usepackage{hyphenat}
\usepackage{hyperref}
\usepackage{placeins}
\usepackage{enumitem}
\usepackage{feynmp-auto}
\usepackage{multirow}
\setcitestyle{numbers,square}

\begin{document}
\title{Wiedemann-Franz law in a non-Fermi liquid and Majorana central charge: \\ Thermoelectric transport in a two-channel Kondo system}
\author{Gerwin A. R. van Dalum}
\affiliation{Institute for Theoretical Physics, Utrecht University, Princetonplein 5, 3584 CC Utrecht, Netherlands}
\author{Andrew K. Mitchell}
\affiliation{School of Physics, University College Dublin, Belfield, Dublin 4, Ireland}
\author{Lars Fritz}
\affiliation{Institute for Theoretical Physics, Utrecht University, Princetonplein 5, 3584 CC Utrecht, Netherlands}


\begin{abstract}
\noindent Quantum dot devices allow one to access the critical point of the two-channel Kondo model. The effective critical theory involves a free Majorana fermion quasiparticle localized on the dot. As a consequence, this critical point shows both the phenomenon of non-Fermi liquid physics and fractionalization. Although a violation of the Wiedemann-Franz law is often considered to be a sign of non-Fermi liquid systems, we show by exact calculations that it holds at the critical point, thereby providing a counterexample to this lore. Furthermore, we show that the fractionalized Majorana character of the critical point can be unambiguously detected from the heat conductance, opening the door to a direct experimental measurement of the elusive Majorana central charge $c=\tfrac{1}{2}$. 
\end{abstract}
\maketitle


Originally conceived to understand the behavior of magnetic impurities in metals~\cite{hewson1997kondo}, the single-channel Kondo model also successfully describes simple quantum dot devices~\cite{goldhaber1998kondo,cronenwett1998tunable,pustilnik2004kondo} and their low-energy Fermi liquid (FL) behavior~\cite{nozieres1974fermi}. Such circuit realizations of fundamental quantum impurity models are exquisitely tunable and allow the nontrivial dynamics of strongly correlated electron systems to be probed experimentally through quantum transport. The FL properties of such systems, as well as their bulk counterparts, are evidenced by their low-temperature thermoelectric transport, which satisfies the Wiedemann-Franz (WF) law~\cite{franz1853wf,costi2010wf}. Conversely, violations of the WF law are observed in various systems with non-Fermi liquid (NFL) properties~\cite{PhysRevLett.95.176602,PhysRevLett.103.096402,PhysRevB.88.125107,_itko_2013,zaanen_liu_sun_schalm_2015,PhysRevB.97.085403,PhysRevB.99.085104,kubala2008violation,dutta2017thermal,crossno2016wf}.

Another advantage of nanoelectronics devices incorporating quantum dots is that more exotic states of quantum matter can be engineered. In particular, there has been considerable interest recently, from both theory and experiment, in \emph{multichannel} Kondo systems~\cite{nozieres1980twochannel} which exhibit NFL quantum critical physics due to frustrated Kondo screening, and the emergence of non-Abelian anyonic quasiparticles~\cite{affleck1991nfl,affleck1993exact,lopes2019anyons}.
The NFL two-channel Kondo (2CK) critical point, realized experimentally in Refs.~\cite{potok2007observation,keller2015universal,iftikhar2015exp}, is described by an effective theory involving Majorana fermions~\cite{emery1992mapping}, while the three-channel Kondo (3CK) critical point realized in Ref.~\cite{iftikhar2018three} involves Fibonacci anyons~\cite{lopes2019anyons}. This NFL character and the fractionalization is most clearly seen in the dot entropy of $S=k_{\text{B}}\ln(\sqrt{2})$ for 2CK and $k_{\text{B}}\ln(\phi)$ for 3CK (with $\phi$ the golden ratio). However, the experimental quantity measured up until now in 2CK and 3CK devices has been the charge conductance~\cite{pustilnik2004quantum,potok2007observation,sela2011exact,*mitchell2012universal,keller2015universal}. In particular, recent charge-Kondo implementations demonstrate precise quantitative agreement between theoretical predictions and experimental measurements for the entire universal scaling curves~\cite{matveev1995setup1,*matveev1995setup2,iftikhar2015exp,mitchell2016universality,iftikhar2018three}. This confirms the underlying theoretical description, but as yet there is no \emph{direct} experimental evidence of either the NFL character or the fractionalization in these systems.

\begin{figure}[t]
\includegraphics[width=0.45\textwidth]{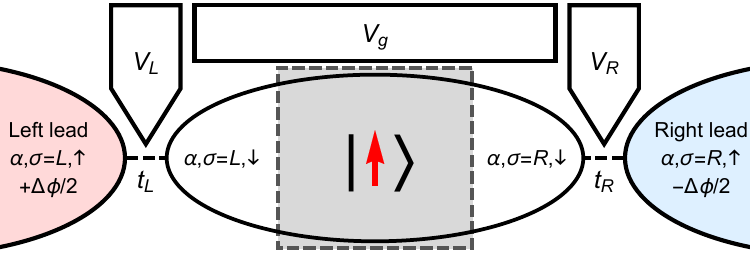}
\caption{Schematic of the C2CK device. Gate voltages $V_{L,R}$ govern the transmission coefficients $t_{L,R}$ at the left and right quantum point contacts, while $V_g$ controls the dot charge. Coherent transport across the dot is suppressed by the intervening ohmic contact. Thermoelectric transport is measured in response to a potential difference $\Delta \phi$ (voltage bias or temperature gradient). A spinless system at the dot charge degeneracy point maps to a 2CK model.}
  \label{fig:schematic}
\end{figure}

Thermoelectric transport in multichannel Kondo systems is far less well understood. In this Rapid Communication, we present exact analytic results for heat transport in the charge-2CK (C2CK) setup depicted in Fig.~\ref{fig:schematic}, relevant to recent experiments~\cite{iftikhar2015exp}. Our choice of system is motivated by the unprecedented control in such a device to probe the NFL critical point; our theoretical predictions are within reach of existing experiments. The C2CK setup allows the WF law~\cite{franz1853wf} to be studied at an exactly solvable NFL critical point. A violation of the WF law has often been used as an empirical rule of thumb to identify NFL physics~\cite{PhysRevLett.95.176602,PhysRevLett.103.096402,PhysRevB.88.125107,_itko_2013,zaanen_liu_sun_schalm_2015,PhysRevB.97.085403,PhysRevB.99.085104}. Nevertheless, we explicitly find that it is satisfied at the charge-2CK NFL critical point. Furthermore, as shown below, the heat conductance is a universal quantity in the critical C2CK system (unlike in the standard spin-2CK implementation), and provides a route to measure experimentally the Majorana central charge. 

We emphasize that we study the non-perturbative regime where both source and drain leads are strongly coupled to the dot (although we focus on linear response corresponding to a small voltage bias and temperature gradient). For this setup, the numerical renormalization group~\cite{bulla2008numerical,*weichselbaum2007sum} (usually considered to be the numerical method of choice for solving generalized quantum impurity problems) cannot be used to calculate heat transport.


\emph{Charge-2CK setup, model, and observables.--}
Fig.~\ref{fig:schematic} shows schematically the C2CK system studied experimentally in Ref.~\cite{iftikhar2015exp}. Ref.~\cite{mitchell2016universality} demonstrated that this quantum dot device realizes an essentially perfect experimental quantum simulation of the C2CK model of Matveev~\cite{matveev1995setup1,*matveev1995setup2}, by comparing experimental data for charge conductance with numerical renormalization group calculations. Here, we compute exact thermoelectric transport analytically at the 2CK critical point for the same model. 

The key ingredient required to realize 2CK physics is ensuring that the two leads constitute two distinct, independent channels (not mixed by inter-channel charge transfer). This is achieved in the C2CK device by exploiting a mapping between charge and (pseudo)spin states~\cite{matveev1995setup1,*matveev1995setup2}. The physical system is effectively spinless (due to the application of a large polarizing magnetic field), and a large dot is tuned to a step in its Coulomb blockade staircase (using gate voltage $V_g$), such that dot charge states with $N$ and $N+1$ electrons are degenerate. Regarding this pair of macroscopic dot charge states as a pseudospin (such that $\hat{S}^{+}=|N+1\rangle\langle N|$ and $\hat{S}^{-}=(S^{+})^{\dagger}$) and simply relabeling dot electrons as ``down'' spin, and lead electrons as ``up'' spin, yields a 2CK pseudospin model -- provided there is no coherent transport between electronic systems around each quantum point contact (QPC). In practice, this is achieved by placing an ohmic contact (metallic island) on the dot to separate the channels~\cite{iftikhar2015exp} (gray box in Fig.~\ref{fig:schematic}). The resulting C2CK Hamiltonian reads
\begin{align}\label{eq:orHam}
H_{\text{K}}&=\sum_{\alpha=L,R} \Bigg [ \sum_{k} (\epsilon_{\alpha \uparrow k}^{\phantom{\dagger}} c_{\alpha\uparrow k}^{\dagger}c_{\alpha \uparrow k}^{\phantom{\dagger}} + \epsilon_{\alpha \downarrow k}^{\phantom{\dagger}} c_{\alpha \downarrow k}^{\dagger}c_{\alpha \downarrow k}^{\phantom{\dagger}}) \nonumber \\
&\!\!\!+t_{\alpha}\sum_{k,k'}( c_{\alpha \uparrow k}^{\dagger}c_{\alpha \downarrow k'}^{\phantom{\dagger}} \hat{S}^{-} + c_{\alpha \downarrow k}^{\dagger}c_{\alpha \uparrow k'}^{\phantom{\dagger}} \hat{S}^{+})\Bigg]+\Delta E \hat{S}^z , 
\end{align}
where $c_{\alpha \sigma k}$ are electronic operators, and $\alpha=L,R$ denotes whether the electron resides to the left or right of the gray metallic island in Fig.~\ref{fig:schematic}. The label $\sigma$ describes whether the electron lives in the leads ($\uparrow$) or on the dot ($\downarrow$). 
The effective continuum of states on the large dot is characterized by the dispersion $\epsilon_{\alpha \downarrow k}$, while the leads have dispersion $\epsilon_{\alpha \uparrow k}$. The term $\Delta E \hat{S}^z$ describes detuning away from the dot charge degeneracy point, which acts as a pseudospin field. Eq.~\eqref{eq:orHam} is a maximally spin-anisotropic version of the regular spin-2CK model~\cite{nozieres1980twochannel}. A major advantage of this setup over the conventional spin-2CK paradigm is that the pseudospin ``exchange'' coupling in the effective model is simply related to the QPC transmission, and can be large. In turn this means that the 2CK Kondo temperature $T_{\text{K}}$ can be high, and hence the critical point is comfortably accessible at experimental base temperatures~\cite{iftikhar2015exp}. The critical point arises for $\Delta E=0$ and $\sqrt{\nu_{L\uparrow}\nu_{L\downarrow}}t_L=\sqrt{\nu_{R\uparrow}\nu_{R\downarrow}}t_R$, where $\nu_{\alpha \sigma}$ is the Fermi level density of states of channel $\alpha \sigma$ (in turn related to the dispersions $\epsilon_{\alpha\sigma k}$). This condition can be achieved~\cite{iftikhar2015exp} by tuning the gate voltages $V_g$, $V_L$ and $V_R$ (Fig.~\ref{fig:schematic}).

We now consider applying a voltage bias $\Delta V$, and/or temperature gradient $\Delta T$, between the left and right leads. The thermoelectric transport coefficients are determined from the resulting charge current $I_c$ and heat current $I_Q$,
\begin{equation}
\label{eq:coeff}
\begin{pmatrix}I_c\\I_Q\end{pmatrix}=\begin{pmatrix}\chi_{cc} & \chi_{cQ}\\ \chi_{Qc} & \chi_{QQ}\end{pmatrix}\begin{pmatrix}\Delta V\\ \Delta T/T\end{pmatrix},
\end{equation}
where $I_{c,Q}\equiv \langle \hat{I}_{c,Q}\rangle$. The current operators are given by
\begin{eqnarray}
\label{eq:currentop}
	\hat{I}_c&=&\frac{e}{2}\frac{\mathrm{d}}{\mathrm{d}t}\left(N_{L,\uparrow}-N_{R,\uparrow}\right)=-\frac{ie}{2\hbar}[N_{L,\uparrow}-N_{R,\uparrow},H] ,\nonumber \\ \hat{I}_Q&=&\frac{i}{2\hbar}\left[H_{L,\uparrow}-H_{R,\uparrow}-\mu\left(N_{L,\uparrow}-N_{R,\uparrow}\right),H \right] .
\end{eqnarray}
With $I_c=G \Delta V$ defined at $\Delta T=0$, and $I_Q=\kappa \Delta T$ defined at $I_c=0$, we
wish to calculate the charge conductance $G=\chi_{cc}$ and heat conductance $\kappa=\left(\chi_{QQ}-\chi_{Qc}\chi_{cQ}/\chi_{cc}\right)/T$.

The charge and heat conductances in linear response can be obtained from the Kubo formula in terms of equilibrium current-current correlation functions~\cite{kubo1957linres,luttinger1964kubo},
\begin{equation}
\label{eq:kubo}
    \chi_{ij} =  \lim_{\omega \to 0}\frac{-{\rm{Im}}\;K_{ij}(\omega,T) }{\hbar\omega} ,
\end{equation}
where $i,j=c,Q$, and $K_{ij}(\omega,T)$ is the Fourier transform of the retarded autocorrelator $K_{ij}(t,T)= -i\theta(t)\langle [\hat{I}_i(t), \hat{I}_j(0) ] \rangle$.


\emph{Emery-Kivelson effective model.--} 
A generalized version of the 2CK model can be solved exactly at a special point in its parameter space, corresponding to a specific value of the exchange anisotropy~\cite{emery1992mapping}. The C2CK model Eq.~\eqref{eq:orHam} (as well as the regular spin-2CK model) does not satisfy this condition. The complete renormalization group (RG) flow and full conductance line shapes at this Emery-Kivelson (EK) point are therefore different from those of the physical system of interest. However, spin anisotropy is RG irrelevant in the 2CK model~\cite{affleck1992relevance,affleck1991nfl,affleck1993exact}, meaning that the same spin-isotropic critical point is reached asymptotically at low temperatures, independently of any anisotropy in the bare model. The EK solution can therefore be used to understand the NFL critical fixed point of the C2CK system~\cite{SM,sengupta1994ek}. This approach has been validated for the entire NFL to FL crossover arising due to small symmetry-breaking perturbations in Refs.~\cite{sela2011exact,*mitchell2012universal,mitchell2016universality} and we adopt the same strategy.

After bosonization, canonical transformation, and refermionization, the EK effective model reads~\cite{emery1992mapping}
\begin{eqnarray}
\label{eq:Ham}
H&=&\sum_\nu\sum_k\epsilon_k\psi^\dagger_{\nu,k}\psi_{\nu,k}+g_\bot[\psi^\dagger_{sf}(0)+\psi_{sf}(0)](d^\dagger-d)\nonumber \\ &+&\frac{\Delta E}{2}(d^\dagger d-dd^\dagger),\label{eq:HEK}
\end{eqnarray} 
where $\psi_{\nu,k}$ (with $\nu=c,s,f,sf$) are effective lead fermion fields, and the impurity spin is parametrized by a fermionic operator $d=i\hat{S}^+$. For all further calculations, we set $\epsilon_k=\hbar v_F k$ for the full range of $k$, where $v_F$ is the Fermi velocity. As a result of the mapping, the effective model takes the form of a non-interacting Majorana resonant level at the critical point $\Delta E=0$. We introduce Majorana operators, 
\begin{eqnarray}
\label{eq:majorana}
\hat{a}=(d^\dagger+d)/\sqrt{2} \quad {\rm{and}}\quad \hat{b}=(d^\dagger-d)/i\sqrt{2},
\end{eqnarray}
such that $\{\hat{a},\hat{a}\}=\{\hat{b},\hat{b}\}=1$ and $\{\hat{a},\hat{b}\}=0$. The effective theory, Eq.~\eqref{eq:Ham}, successfully accounts for the residual fractional dot entropy at the C2CK critical point, arising from the strictly decoupled $\hat{a}$ Majorana.

 A remarkable feature of the EK mapping is that it holds even with a finite voltage bias between leads, allowing charge transport to be calculated beyond linear response~\cite{schiller1998toulouse}. However, this approach cannot be used for non-equilibrium transport in the presence of a temperature difference between the leads because the EK mapping mixes the two electronic baths, and so the EK channels cannot be assigned a definite temperature. In the following, we therefore confine attention to thermoelectric transport in linear response using the Kubo formula~\cite{kubo1957linres}.


\emph{Current operators in the EK basis.--} Transforming the charge and heat current operators, Eq.~\eqref{eq:currentop}, into the EK basis and writing in terms of dot Majorana operators $\hat{a}$ and $\hat{b}$ from Eq.~\eqref{eq:majorana}, we find
\begin{eqnarray}
\label{eq:IC}
\hat{I}_c=\frac{eg_\bot}{\sqrt{2L}\hbar}\sum_k\left(\psi^\dagger_{sf,k}-\psi_{sf,k}\right)\hat{b} ,\label{eq:Ic}
\end{eqnarray}
for the charge current, but for the heat current,
\begin{eqnarray}\label{eq:IE}
&\hat{I}_Q&=\frac{i\pi v_Fg_\bot}{(2L)^{3/2}}\sum\limits_{\mathclap{k,k^\prime,k^{\prime\prime}}}\left(2\,\psi_{f,k^\prime}^\dagger\psi_{f,k^{\prime\prime}}+\delta_{k^\prime,k^{\prime\prime}}\right)\left(\psi_{sf,k}^\dagger+\psi_{sf,k}\right)\hat{a}\nonumber \\ &-&\frac{\pi v_Fg_\bot}{\sqrt{2}L^{3/2}}\sum\limits_{\mathclap{k,k^\prime,k^{\prime\prime}}}\left(\psi_{c,k^\prime}^\dagger\psi_{c,k^{\prime\prime}}+\psi_{s,k^\prime}^\dagger\psi_{s,k^{\prime\prime}}\right)\left(\psi_{sf,k}^\dagger-\psi_{sf,k}\right)\hat{b}\nonumber\\
&+&\frac{\pi v_F}{2L}\sum\limits_{\mathclap{k,k^\prime}}\left(\epsilon_{k^\prime}-\epsilon_k\right)\left(\psi_{f,k}^\dagger\psi_{f,k^\prime}+\psi_{sf,k}^\dagger\psi_{sf,k^\prime}\right)\hat{a}\hat{b} .
\end{eqnarray}
Here, $L$ is the length of a lead, and we have set $\mu=0$.

\begin{figure}[t]
\includegraphics[width=0.45\textwidth]{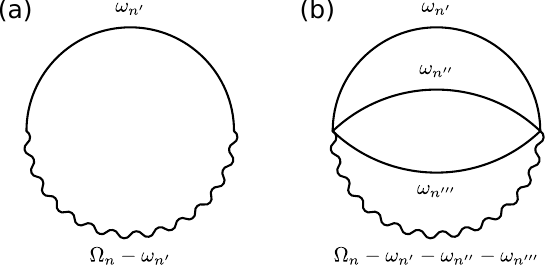}
\caption{The only Feynman diagrams contributing to (a) the charge conductance $G$, and (b) the heat conductance $\kappa$. $\Omega_n$ represents the external bosonic Matsubara frequency, and we sum over the remaining fermionic Matsubara frequencies $\omega_n$.}
  \label{fig:diagrams}
\end{figure}

\emph{Linear response coefficients.--} The calculation of the charge conductance $G$ is rather straightforward~\cite{schiller1998toulouse}, involving as it does only one-loop Feynman diagrams of the type shown in Fig.~\ref{fig:diagrams}(a). Here, we represent diagrammatically the local (imaginary time) bare bath propagators $L^0_\nu(\tau)=-1/L \sum_k\langle \hat{T} \psi^{\phantom{\dagger}}_{\nu,k}(\tau) \psi^\dagger_{\nu,k} (0) \rangle_0$ (with $\nu=c,s,f,sf$) using  ``straight'' lines, while the fully renormalized Majorana Green function $D_{bb}(\tau)=\langle \hat{T} b(0)b(\tau) \rangle$ is represented diagrammatically as a ``wiggly'' line. For more details on the calculation and the definition of the Green function, see the Supplemental Material~\cite{SM}.

At the critical point, the EK calculation yields the well-known leading order in temperature result for the charge conductance~\cite{schiller1998toulouse,mitchell2016universality},
  \begin{eqnarray}\label{eq:cond}
  G=\frac{e^2}{2h}.
  \end{eqnarray}

By contrast, the heat conductance calculation is far more involved. In this case, one must compute three-loop Feynman diagrams of the type shown in Fig.~\ref{fig:diagrams}(b). After a lengthy calculation~\cite{SM}, we find the following form for the leading order low-temperature heat conductance, 
 \begin{eqnarray}\label{eq:main}
\kappa=\frac{\pi^2 k_B^2 T}{6h},
\end{eqnarray}
and the off-diagonal components $\chi_{cQ}=\chi_{Qc}$ vanish. These are exact results at the critical point of the C2CK system. Eq.~\eqref{eq:main} is our central result, the physical consequences of which are explored in detail in the following.

\emph{Applicability of the EK solution.--} The leading order finite-temperature corrections to Eqs.~\eqref{eq:cond} and \eqref{eq:main} are linear in $T$. They originate from the leading irrelevant operator, of scaling dimension 3/2, which is $H_I=\frac{i\lambda}{L}\hat{b}\hat{a}\sum_{k,k^\prime}:\psi_{s,k}^\dagger\psi_{s,k^\prime}:$~\cite{emery1992mapping}, corresponding to spin anisotropy. In the Supplemental Material~\cite{SM} we show that this implies $G=\frac{e^2}{2h}\left(1-\frac{\pi^3\lambda^2}{8h^2v_F^2}\frac{T}{T_K}+\ldots\right)$ where $T_K$ is the Kondo temperature. A similar calculation for the heat conductance is five-loop, which we did not attempt. However, the structure of the perturbation theory implies a similar generic form,  $\kappa=\frac{\pi^2 k_B^2 T}{6h}\left(1-b \lambda^2 \frac{T}{T_K}+\ldots\right)$. Both $G$ and $\kappa$ are \emph{finite} at the EK point, with leading corrections controlled by powers of $T/T_K$ which vanish at the critical C2CK fixed point as $T/T_K\rightarrow 0$.

\emph{Wiedemann-Franz law.--}
For weakly interacting metals, Wiedemann and Franz found~\cite{franz1853wf} a remarkable relation between the low-temperature electrical and thermal conductivities: 
$\lim_{T/T_F \to 0}\kappa/(T\sigma) = L_0$, where $L_0=\pi^2k_\text{B}^2/3e^2$ is the Lorenz number which involves only fundamental constants and $T_F$ is the Fermi temperature (the relation is asymptotic, based on a leading order expansion in $T/T_F$). Metals are good conductors of both charge and heat, since the carriers in both cases are itinerant electrons. Such a relation also holds in the context of many nanoelectronics systems at low temperatures (where the conductance $G$ plays the role of $\sigma$), even with strong electronic correlations~\cite{costi2010wf} -- provided the system is a Fermi liquid at low temperatures. Indeed, a violation of the WF law is often considered a hallmark of non-Fermi liquid physics, since there the carriers are not simply bare electrons or ``dressed'' fermionic quasiparticles as in FL theory, but more complicated objects, possibly with different and even fractionalized quantum numbers.

The C2CK system offers a rare opportunity to test the WF law at an exactly solvable NFL critical point, and to make a concrete prediction for experiments. Interestingly -- and contrary to conventional expectation -- we find that the WF law is \emph{satisfied} at the C2CK NFL critical point,
\begin{eqnarray}\label{WFexact}
\lim_{T/T_K\to 0}\frac{\kappa}{T G}=	\frac{\pi^2k_\text{B}^2}{3e^2}.
\end{eqnarray}
Since $G$ and $\kappa$ are both finite at the C2CK critical fixed point, and corrections to the fixed point are strictly RG irrelevant, Eq.~(\ref{WFexact}) is exact.

The WF law is expected to be \emph{violated} in the FL phase of the C2CK model. After preparing the C2CK system at the NFL critical point, consider introducing a small symmetry-breaking perturbation (coupling the dot more strongly to one lead than the other). The system flows under RG on further reducing temperature to a Fermi liquid state, in which the dot pseudospin is fully Kondo screened by one lead, while the other asymptotically decouples~\cite{matveev1995setup1,*matveev1995setup2}. The resulting charge conductance $G\rightarrow 0$ in the FL phase~\cite{mitchell2016universality}, since one of the two physical leads involved in transport decouples. For the same reasons, $\kappa/T\rightarrow 0$. The WF ratio, obtained in the limiting process of $T/T_K\rightarrow 0$, is therefore expected to take a non-universal value with a different Lorenz ratio $L\ne L_0$ due to the leading temperature-dependent corrections to the FL fixed point values of $G$ and $\kappa$. 


\emph{Measuring the Majorana central charge.--} 
 The EK effective model for the C2CK system is essentially a one-dimensional (1D) boundary problem.  Critical systems in 1D are described by conformal field theories in $1+1$ dimensions, as characterized by the so-called conformal charge $c$. Recently, it was conjectured that heat transport is directly proportional to the conformal charge of the underlying conformal field theory~\cite{bernard2012central}. For translationally invariant critical systems with left and right leads held at temperatures $T_L$ and $T_R$, the heat current is given by $I_Q=\pi^2 k_{\text{B}}^2 c (T_L^2-T_R^2)/6h$. Within linear response, we take $T_R=T$ and $T_L=T+\Delta T$, and expand to leading order in $\Delta T$,
\begin{eqnarray}
I_Q = \frac{\pi^2 k_{\text{B}}^2}{3h}c T \Delta T +\mathcal{O}\left[(\Delta T)^2\right] ,
\end{eqnarray}
which allows us to identify $\kappa$ as 
\begin{eqnarray}
\kappa =	\frac{\pi^2 k_{\text{B}}^2}{3h}c T.
\end{eqnarray}
Comparing this to Eq.~\eqref{eq:main}, we find that the central charge of the underlying effective critical theory is $c=\tfrac{1}{2}$. This is consistent with the known result of $c=\tfrac{1}{2}$ for one-dimensional Majorana fermions in the unitary limit~\cite{affleck1993exact}. We argue that heat transport measurements therefore provide clear experimental access to the fractionalized nature of the excitations in the C2CK system. 


\emph{Comparison with spin-2CK.--} 
The above results are specific to the C2CK setup relevant to recent experiments~\cite{iftikhar2015exp,iftikhar2018three}. Here, we briefly contrast to the standard spin-2CK setup of Refs.~\cite{potok2007observation,keller2015universal}, in which one of the two conduction electron channels is ``split'' into source and drain leads, with their hybridization to the quantum dot parametrized by $\Gamma_s$ and $\Gamma_d$, respectively (the other channel is a Coulomb blockaded quantum box). Although the effective EK model at the 2CK critical point is the same, the form of the current operators [the analog of Eqs.~\eqref{eq:IC} and \eqref{eq:IE}] is obviously different. Indeed, the ``proportionate coupling'' geometry of that setup affords a significant simplification, with charge and heat conductances expressible simply in terms of the scattering $t$-matrix spectrum $t(T,\omega)$ as shown in Ref.~\cite{costi2010wf}. 
At the spin-2CK critical fixed point, the charge conductance for $T/T_K\rightarrow 0$ follows as $G=2\gamma e^2 t(0,0)/h$, while the heat conductance is $\kappa= 2\gamma\pi^2 k_{\text{B}}^2T t(0,0)/3h$, where the geometrical factor is $\gamma=4\Gamma_s\Gamma_d/(\Gamma_s+\Gamma_d)^2$, and at the 2CK fixed point we have $t(0,0)=\tfrac{1}{2}$~\cite{affleck1993exact,sela2011exact,*mitchell2012universal}. The spin-2CK conductances are therefore not universal and depend on system geometry through $\gamma$. There is no interpretation in terms of the central charge since the setup is not a translationally invariant 1D system. On the other hand, the WF law is satisfied since $L=\kappa/TG=\pi^2 k_{\text{B}}^2/3e^2=L_0$. In this setup, channel asymmetry produces a flow away from the NFL critical point and towards a low-temperature FL state, in which the leads probing transport can be either in strong coupling (SC) or weak coupling (WC) with the dot (depending on which channel couples more strongly). At SC, $t(0,0)=1$ and the WF law is again satisfied. However, at WC, $t(0,0)=0$ and the leading (quadratic) Fermi liquid corrections to the $t$-matrix must be considered~\cite{sela2011exact,*mitchell2012universal}. In this case we find a different universal ratio $L=7\pi^2 k_{\text{B}}^2/5e^2\ne L_0$. A similar analysis can be performed for the spin-3CK situation~\cite{mitchell2014generalized,*stadler2016interleaved}, where the NFL fixed point is characterized by $t(0,0)=\cos(2\pi/5)$.

\emph{Failure of NRG for heat transport via Kubo.--} 
Finally, we comment that our exact analytic results for heat conductance in the C2CK system are, perhaps surprisingly, \emph{inaccessible} with the numerical renormalization group (NRG)~\cite{bulla2008numerical,*weichselbaum2007sum}.
If a system satisfies ``proportionate coupling'', thermoelectric transport coefficients may be related to moments of the scattering $t$-matrix, and NRG can be used to obtain accurate results, as demonstrated in Refs.~\cite{costi1994nrg,costi2010wf} for the Anderson model. However, the geometry of the setup depicted in Fig.~\ref{fig:schematic} does not admit any such formulation of the conductances in terms of the $t$-matrix, and one must fall back on the Kubo formula, Eq.~(\ref{eq:kubo}). 
The latter uses the heat current operator Eq.~\eqref{eq:currentop}, which involves the lead Hamiltonian $H_{\alpha,\uparrow}$. In NRG, a specific discretized form of $H_{\alpha,\uparrow}$ is used, but these ``Wilson chains'' do not act as proper thermal reservoirs~\cite{rosch2012wilson}. We find that exact FL results for even the simple resonant level model cannot be reproduced with the Kubo formula when Wilson chains are used for leads. NRG can of course be used to compute impurity dynamical quantities or the $t$-matrix~\cite{bulla2008numerical,*weichselbaum2007sum}, and the Kubo formula may still be used for charge transport within NRG~\cite{mitchell2016universality,iftikhar2018three,mitchell2017kondo,galpin2014conductance}.


\emph{Conclusions and outlook.--} 
We studied charge and heat transport in the C2CK system recently realized experimentally~\cite{iftikhar2015exp}, by exploiting the exact solution of the related EK model~\cite{emery1992mapping} and RG arguments. In particular, our new result for the low-temperature heat conductance at the NFL critical point, $\kappa=\pi^2k_{\text{B}}^2T/6h$, is exact. Our results show that the WF law is satisfied, despite being a NFL. Furthermore, we demonstrate that the heat transport provides an experimental route to determine the central charge of the underlying conformal field theory, which in this case is $c=\tfrac{1}{2}$ because an effective Majorana fermion mediates charge and heat transport through the dot. It would be interesting to extend this study to the charge-3CK system in the regime where all leads are coupled non-perturbatively. This is a formidable theoretical challenge since there is no equivalent exact solution available as with C2CK, and one should expect WF to be violated. We note that heat transport measurements in a C3CK system are within existing experimental reach~\cite{iftikhar2018three}.\vspace{0.09cm}


\emph{Note added.--} 
Recently, we became aware of Ref.~\cite{nguyen2019thermoelectric}, which considers the closely related problem of thermoelectric transport in a three-channel charge Kondo problem with an additional weakly coupled probe lead.

\emph{Acknowledgement.--} We thank P. Simon and E. Sela for helpful discussions. L.F. and G.D. acknowledge funding from the D-ITP consortium, a program of the Netherlands Organisation for Scientific Research (NWO) that is funded by the Dutch Ministry of Education, Culture and Science (OCW). A.K.M. acknowledges funding from the Irish
Research Council Laureate Awards 2017/2018 through grant IRCLA/2017/169.

%

\onecolumngrid

\clearpage

\begin{center}{\large\textbf{Supplemental Material for ``Wiedemann-Franz law in a non-Fermi liquid and Majorana central charge: Thermoelectric transport in a two-channel Kondo system''}}\\[0.5cm]
Gerwin A. R. van Dalum\textsuperscript{1}, Andrew K. Mitchell\textsuperscript{2}, and Lars Fritz\textsuperscript{1}\\[0.2cm]
\textsuperscript{1}\textit{Institute for Theoretical Physics, Utrecht University, Princetonplein 5, 3584 CC Utrecht, Netherlands}\\
\textsuperscript{2}\textit{School of Physics, University College Dublin, Belfield, Dublin 4, Ireland}
\end{center}

\renewcommand{\theequation}{S\arabic{equation}}
\renewcommand{\thefigure}{S\arabic{figure}}
\setcounter{figure}{0}
\setcounter{equation}{0}

\section*{Emery-Kivelson mapping of the current operators}
We show how the heat current operator with $\mu=0$ is given by Eq.~(\ref{eq:IE}) in the EK basis. The first part of the mapping procedure~\cite{emery1992mapping} is the introduction of a bosonic field $\Phi_{\alpha\sigma}(x)$ for each of the fermionic fields $c_{\alpha\sigma}(x)$:
\begin{equation}
c_{\alpha\sigma}(x)=\frac{1}{\sqrt{2\pi a_0}}e^{i\phi_{\alpha\sigma}}e^{-i\Phi_{\alpha\sigma}(x)};
\end{equation}
the exponentials $e^{i\phi_{\alpha\sigma}}$ act as Klein factors to ensure the correct anticommutation relations between the fermionic fields. Following the usual bosonization prescription, the various components of the ``charge'' operators $\hat{Q}_c=-e(N_{L,\uparrow}-N_{R,\uparrow})/2$ and $\hat{Q}_E=(H_{L,\uparrow}-H_{R,\uparrow})/2$ transform according to
\begin{align}
\int\limits_{-\infty}^{\infty}\mathrm{d}x\,c^\dagger_{\alpha\sigma}(x)c_{\alpha\sigma}(x)&=\frac{1}{2\pi}\int\limits_{-\infty}^{\infty}\mathrm{d}x\,\partial_x\Phi_{\alpha\sigma}(x),\label{eq:bosonization1}\\
\int\limits_{-\infty}^{\infty}\mathrm{d}x\,c^\dagger_{\alpha\sigma}(x)\partial_xc_{\alpha\sigma}(x)&=-\frac{i}{4\pi}\int\limits_{-\infty}^{\infty}\mathrm{d}x\left(\partial_x\Phi_{\alpha\sigma}(x)\right)^2,\label{eq:bosonization2}
\end{align}
where normal ordering of the fermionic fields is implied. In terms of the linear combinations
\begin{align}
\Phi_c(x)&\equiv\frac{1}{2}\big(\Phi_{L\uparrow}(x)+\Phi_{L\downarrow}(x)+\Phi_{R\uparrow}(x)+\Phi_{R\downarrow}(x)\big),\label{eq:chargemode}\\
\Phi_s(x)&\equiv\frac{1}{2}\big(\Phi_{L\uparrow}(x)-\Phi_{L\downarrow}(x)+\Phi_{R\uparrow}(x)-\Phi_{R\downarrow}(x)\big),\\
\Phi_f(x)&\equiv\frac{1}{2}\big(\Phi_{L\uparrow}(x)+\Phi_{L\downarrow}(x)-\Phi_{R\uparrow}(x)-\Phi_{R\downarrow}(x)\big),\label{eq:flavor}\\
\Phi_{sf}(x)&\equiv\frac{1}{2}\big(\Phi_{L\uparrow}(x)-\Phi_{L\downarrow}(x)-\Phi_{R\uparrow}(x)+\Phi_{R\downarrow}(x)\big),\label{eq:spinflavor}
\end{align}
this leads to
\begin{align}
\hat{Q}_c&=-\frac{e}{4\pi}\int\limits_{-\infty}^{\infty}\mathrm{d}x\big(\partial_x\Phi_f(x)+\partial_x\Phi_{sf}(x)\big),\\
\hat{Q}_E&=\frac{\hbar v_F}{8\pi}\int\limits_{-\infty}^{\infty}\mathrm{d}x\big(\partial_x\Phi_c(x)+\partial_x\Phi_s(x)\big)\big(\partial_x\Phi_f(x)+\partial_x\Phi_{sf}(x)\big).
\end{align}
The next step of the EK mapping procedure is the unitary transformation $\hat{\mathcal{O}}\rightarrow\hat{U}\hat{\mathcal{O}}\hat{U}^\dagger$, with $\hat{U}=e^{i\chi_s\hat{S}^z}$ and $\chi_s\equiv\Phi_s(0)-\phi_s$. Using the commutation relation
\begin{equation}
\left[\Phi_\mu(x),\partial_x\Phi_\nu(x^\prime)\right]=2\pi i\,\delta_{\mu,\nu}\,\delta(x-x^\prime)\label{eq:commutation}
\end{equation}
together with $d\equiv i\hat{S}^+$ (such that $\hat{S}^z=-(d^\dagger d-1/2)$), it is straightforward to show that
\begin{equation}
\hat{Q}_E\rightarrow\frac{\hbar v_F}{8\pi}\int\limits_{-\infty}^{\infty}\mathrm{d}x\big(\partial_x\Phi_c(x)+\partial_x\Phi_s(x)\big)\big(\partial_x\Phi_f(x)+\partial_x\Phi_{sf}(x)\big)+\frac{\hbar v_F}{4}\left(d^\dagger d-\frac{1}{2}\right)\big(\partial_x\Phi_f(x)+\partial_x\Phi_{sf}(x)\big)\Big|_{x=0}
\end{equation}
under this unitary transformation, while $\hat{Q}_c$ remains unchanged. The final step of the mapping procedure consists of refermionization. Using relations similar to those involved in the initial bosonization step and noting that
\begin{equation}
\int\limits_{-\infty}^{\infty}\mathrm{d}x\,\psi^\dagger_\mu(x)\psi_\mu(x)\psi^\dagger_\nu(x)\psi_\nu(x)=\frac{1}{4\pi^2}\int\limits_{-\infty}^{\infty}\mathrm{d}x\big(\partial_x\Phi_\mu(x)\big)\big(\partial_x\Phi_\nu(x)\big)\label{eq:bosonization3}
\end{equation}
for $\mu\neq\nu$ (normal ordering of the fermionic fields again implied), we find that the charge operators finally become
\begin{align}
\hat{Q}_c&=-\frac{e}{2}\int\limits_{-\infty}^{\infty}\mathrm{d}x\Big(\psi^\dagger_f(x)\psi_f(x)+\psi^\dagger_{sf}(x)\psi_{sf}(x)\Big),\\
\hat{Q}_E&=\frac{\pi\hbar v_F}{2}\int\limits_{-\infty}^{\infty}\mathrm{d}x\Big(\psi_c^\dagger(x)\psi_c(x)+\psi_s^\dagger(x)\psi_s(x)\Big)\Big(\psi_f^\dagger(x)\psi_f(x)+\psi_{sf}^\dagger(x)\psi_{sf}(x)\Big)\nonumber\\
&\quad\,+\frac{\pi\hbar v_F}{2}\Big(:\psi_f^\dagger(0)\psi_f(0):+:\psi_{sf}^\dagger(0)\psi_{sf}(0):\Big)\left(d^\dagger d-\frac{1}{2}\right).
\end{align}

We now calculate the current operators by Fourier transforming the charge operators to momentum space and evaluating the commutators with the Hamiltonian from Eq.~(\ref{eq:HEK}). Starting with electric transport:
\begin{align}
\hat{I}_c&=-\frac{ie}{2\hbar}\sum_k\left[\psi^\dagger_{f,k}\psi_{f,k}+\psi^\dagger_{sf,k}\psi_{sf,k},\hat{H}\right]\nonumber\\
&=-\frac{ieg_\bot}{2\hbar\sqrt{L}}\sum_k\left(\psi^\dagger_{sf,k}-\psi_{sf,k}\right)\left(d^\dagger-d\right).
\end{align}
Although more cumbersome, the energy current operator can be obtained in the same way, leading to
\begin{align}
\hat{I}_E&=\frac{i\pi v_Fg_\bot}{2L^{3/2}}\sum\limits_{\mathclap{k,k^\prime,k^{\prime\prime}}}\left(\psi_{c,k^\prime}^\dagger\psi_{c,k^{\prime\prime}}+\psi_{s,k^\prime}^\dagger\psi_{s,k^{\prime\prime}}\right)\left(\psi_{sf,k}^\dagger-\psi_{sf,k}\right)\left(d^\dagger-d\right)\nonumber\\
&\quad\,+\frac{i\pi v_Fg_\bot}{4L^{3/2}}\sum\limits_{\mathclap{k,k^\prime,k^{\prime\prime}}}\left(2\,\psi_{f,k^\prime}^\dagger\psi_{f,k^{\prime\prime}}+\delta_{k^\prime,k^{\prime\prime}}\right)\left(\psi_{sf,k}^\dagger+\psi_{sf,k}\right)\left(d^\dagger+d\right)\nonumber\\
&\quad\,+\frac{i\pi v_F}{4L}\sum\limits_{\mathclap{k,k^\prime}}\left(\epsilon_{k^\prime}-\epsilon_k\right)\left(\psi_{f,k}^\dagger\psi_{f,k^\prime}+\psi_{sf,k}^\dagger\psi_{sf,k^\prime}\right)\left(d^\dagger d-dd^\dagger\right).
\end{align}

\section*{Structure of the Green function}
In order to derive the linear response heat conductance at the NFL fixed point, we first consider the propagator structure of the model from Eq.~(\ref{eq:HEK}) with $\Delta E=0$. Only considering the $\nu=sf$ modes for now (since the $\nu=c,s,f$ modes are decoupled), the Green function has the following matrix structure:
\begin{equation}
\mathbf{G}\equiv\begin{pmatrix}\mathbf{L} & \mathbf{G}_{ld} \\ \mathbf{G}_{dl} & \mathbf{D} \end{pmatrix}=\begin{pmatrix}\mathbf{L}_0^{-1} & -\mathbf{g}_\bot/\hbar \\ -\mathbf{g}^\dagger_\bot/\hbar & \mathbf{D}_0^{-1} \end{pmatrix}^{-1},\label{eq:greensinv}
\end{equation}
where $\mathbf{L}$ and $\mathbf{D}$ are the full equilibrium Green functions of the $\nu=sf$ lead modes and the dot, respectively, and the subscript $0$ refers to the bare propagators in absence of tunneling. To incorporate the Majorana nature of the tunneling processes, we switch to the Nambu spinor basis, for example working with $\mathbf{d}^\dagger\equiv(d^\dagger\;d)$. In momentum space, all components of the tunneling matrix (labeled by index $k$) can be deduced from Eq.~(\ref{eq:HEK}), and are given by
\begin{equation}
\mathbf{g}_{\bot,k}=\frac{g_{\bot}}{\sqrt{L}}\begin{pmatrix}-1 & 1 \\ -1 & 1\end{pmatrix}\equiv\frac{g_\bot}{\sqrt{L}}\,\mathbf{g},\label{eq:gbotm}
\end{equation}
independent of $k$. Moreover, all of the momentum space components of the Green functions are $2\times 2$ matrices as well; block inversion of the right-hand side of Eq.~\eqref{eq:greensinv} leads to
\begin{align}
\mathbf{D}&=\left(\mathbf{D}_0^{-1}-\pmb{\Sigma}_d\right)^{-1},\\
\mathbf{G}_{ld,k}&=\frac{g_\bot}{\hbar\sqrt{L}}\,\mathbf{L}_{0,k}\cdot\mathbf{g}\cdot\mathbf{D},\label{eq:Gldk}\\
\mathbf{L}_{kk^\prime}&=\delta_{k,k^\prime}\mathbf{L}_{0,k}+\frac{g_\bot^2}{\hbar^2L}\,\mathbf{L}_{0,k}\cdot\mathbf{g}\cdot\mathbf{D}\cdot\mathbf{g}^\dagger\cdot\mathbf{L}_{0,k^\prime},
\end{align}
with the dot self-energy being equal to
\begin{equation}
\pmb{\Sigma}_d=\frac{g_\bot^2}{\hbar^2}\,\mathbf{g}^\dagger\cdot\Big(\frac{1}{L}\sum_k\mathbf{L}_{0,k}\Big)\cdot\mathbf{g}\equiv\frac{g_\bot^2}{\hbar^2}\,\mathbf{g}^\dagger\cdot\mathbf{L}_0^\prime\cdot\mathbf{g}.\label{eq:sigmad}
\end{equation}
For future reference, we also introduce the Majorana Green functions on the dot, corresponding to the Majorana fermions $a$ and $b$; they are given by
\begin{align}
D_{aa}&=\frac{1}{2}\left(D_{11}+D_{12}+D_{21}+D_{22}\right),\label{eq:majoranagreens1}\\
D_{bb}&=\frac{1}{2}\left(D_{11}-D_{12}-D_{21}+D_{22}\right),\\
D_{ab}&=\frac{1}{2i}\left(D_{11}-D_{12}+D_{21}-D_{22}\right),\\
D_{ba}&=\frac{1}{2i}\left(-D_{11}-D_{12}+D_{21}+D_{22}\right),\label{eq:majoranagreens2}
\end{align}
where $D_{ij}$ are the original components of the $2\times 2$ matrix $\mathbf{D}$. Finally, it should be noted that all of the above fields and Green functions have implied time-dependence.

In terms of fermionic Matsubara frequencies $\omega_n$, the required Green functions are given by
\begin{gather}
\mathbf{L}_{0,k}(i\omega_n)=\hbar\begin{pmatrix}(i\hbar\omega_n-\epsilon_k)^{-1} & 0 \\ 0 & (i\hbar\omega_n+\epsilon_k)^{-1}\end{pmatrix},\label{eq:L0k}\\
\mathbf{D}(i\omega_n)\equiv\mathbf{G}_{dd}(i\omega_n)=\int\limits_{-\infty}^\infty\mathrm{d}\epsilon\,\frac{\pmb{\rho}(\epsilon)}{i\hbar\omega_n-\epsilon},\qquad\pmb{\rho}(\epsilon)\equiv-\frac{1}{\pi}\text{Im}\!\left[\mathbf{D}^\text{R}(\epsilon)\right],\label{eq:Dw}
\end{gather}
where $\pmb{\rho}$ can be interpreted as a density of states, and the retarded dot Green function is given by~\cite{schiller1998toulouse}
\begin{equation}
\mathbf{D}^\text{R}(\epsilon)=\frac{\hbar}{\epsilon(\epsilon+i\Gamma)}\begin{pmatrix}\epsilon+\frac{i}{2}\Gamma & \frac{i}{2}\Gamma \\ \frac{i}{2}\Gamma & \epsilon+\frac{i}{2}\Gamma\end{pmatrix}.\label{eq:DR}
\end{equation}
Here, the parameter $\Gamma$ has been introduced for notational convenience and for later reference; it is defined as
\begin{equation}
\Gamma\equiv 2g_\bot^2\frac{\mathrm{d}k}{\mathrm{d}\epsilon_k}=\frac{2g_\bot^2}{\hbar v_F}.
\end{equation}
We thus find:
\begin{align}
D_{aa}(i\omega_n)&=\frac{1}{i\omega_n},\\
D_{bb}(i\omega_n)&=-\frac{i\hbar}{\hbar\omega_n+\text{sgn}(\omega_n)\Gamma},\label{eq:Dbbomega}\\
D_{ab}(i\omega_n)&=D_{ba}(i\omega_n)=0.
\end{align}
Finally, we use the above to point out that the Green functions satisfy the following equations:
\begin{align}
    \sum_{\mu\nu} G_{dd,\mu\nu}(i\omega_n)&=2D_{aa}(i\omega_n),\label{eq:unsignedsums1}\\
    \sum_{\mu\nu} G_{ld,k,\mu\nu}(i\omega_n)&=\frac{4g_\bot}{\sqrt{L}}\frac{\hbar\omega_n}{(\hbar\omega_n)^2+\epsilon_k^2}D_{ba}(i\omega_n)=0,\label{eq:unsignedsums2}\\
    \sum_{\mu\nu} G_{ll,kk^\prime,\mu\nu}(i\omega_n)&=-2i\hbar\,\delta_{k,k^\prime}\frac{\hbar\omega_n}{(\hbar\omega_n)^2+\epsilon_k^2}-\frac{8g_\bot^2}{L}\frac{\hbar\omega_n}{(\hbar\omega_n)^2+\epsilon_k^2}\frac{\hbar\omega_n}{(\hbar\omega_n)^2+\epsilon_{k^\prime}^2}D_{bb}(i\omega_n),
\end{align}
\begin{align}
\sum_{\mu\nu}^\prime G_{dd,\mu\nu}(i\omega_n)&=2D_{bb}(i\omega_n),\label{eq:signedsums1}\\
\sum_{\mu\nu}^\prime G_{ld,k,\mu\nu}(i\omega_n)&=\frac{4g_\bot}{\sqrt{L}}\frac{\epsilon_k}{(\hbar\omega_n)^2+\epsilon_k^2}D_{bb}(i\omega_n),\label{eq:signedsums2}\\
\sum_{\mu\nu}^\prime G_{ll,kk^\prime,\mu\nu}(i\omega_n)&=-2i\hbar\,\delta_{k,k^\prime}\frac{\hbar\omega_n}{(\hbar\omega_n)^2+\epsilon_k^2}+\frac{8g_\bot^2}{L}\frac{\epsilon_k}{(\hbar\omega_n)^2+\epsilon_k^2}\frac{\epsilon_{k^\prime}}{(\hbar\omega_n)^2+\epsilon_{k^\prime}^2}D_{bb}(i\omega_n),\label{eq:signedsums3}\\
G_{ld,k,11}(i\omega_n)-G&_{ld,k,22}(i\omega_n)-G_{ld,k,12}(i\omega_n)+G_{ld,k,21}(i\omega_n)=\frac{4ig_\bot}{\sqrt{L}}\frac{\hbar\omega_n}{(\hbar\omega_n)^2+\epsilon_k^2}D_{bb}(i\omega_n),\label{eq:unsignedsums4}
\end{align}
where the unprimed sums denote normal sums over all components, and the primed sums are signed sums in which the off-diagonal components $\mu\neq\nu$ pick up a minus sign.

\section*{Calculating the linear heat susceptibility}
In terms of the imaginary time $\tau$, the required autocorrelator is given by
\begin{equation}
K^\tau_{ij}(\tau-\tau^\prime,T)=-\big\langle\hat{T}\hat{I}_i(\tau)\hat{I}_j(\tau^\prime)\big\rangle,
\end{equation}
where $i,j=c,Q$, and $\hat{T}$ is the imaginary time ordering operator. To calculate the heat susceptibility, we first decompose the heat current operator into five separate terms: $\hat{I}_Q=\sum_{i=1}^5\hat{I}_i$, with
\begin{align}
\hat{I}_1&=-\frac{\pi v_Fg_\bot}{\sqrt{2}L^{3/2}}\sum\limits_{\mathclap{k,k^\prime,k^{\prime\prime}}}\left(\psi_{c,k^\prime}^\dagger\psi_{c,k^{\prime\prime}}+\psi_{s,k^\prime}^\dagger\psi_{s,k^{\prime\prime}}\right)\left(\psi_{sf,k}^\dagger-\psi_{sf,k}\right)b,\label{eq:I1}\\
\hat{I}_2&=\frac{i\pi v_Fg_\bot}{\sqrt{2}L^{3/2}}\sum\limits_{\mathclap{k,k^\prime,k^{\prime\prime}}}\psi_{f,k^\prime}^\dagger\psi_{f,k^{\prime\prime}}\left(\psi_{sf,k}^\dagger+\psi_{sf,k}\right)a,\label{eq:I2}\\
\hat{I}_3&=\frac{i\Lambda g_\bot}{2^{3/2}\hbar\sqrt{L}}\sum\limits_{\mathclap{k}}\left(\psi_{sf,k}^\dagger+\psi_{sf,k}\right)a,\label{eq:I3}\\
\hat{I}_4&=\frac{\pi v_F}{2L}\sum\limits_{\mathclap{k,k^\prime}}\left(\epsilon_{k^\prime}-\epsilon_k\right)\psi_{f,k}^\dagger\psi_{f,k^\prime}ab,\label{eq:I4}\\
\hat{I}_5&=\frac{\pi v_F}{2L}\sum\limits_{\mathclap{k,k^\prime}}\left(\epsilon_{k^\prime}-\epsilon_k\right)\psi_{sf,k}^\dagger\psi_{sf,k^\prime}ab.\label{eq:I5}
\end{align}
Here, $\Lambda$ is the energy cut-off that is introduced by writing $\int_{-\infty}^\infty\mathrm{d}\epsilon_k\rightarrow\int_{-\Lambda}^\Lambda\mathrm{d}\epsilon_k$. In addition, it is useful to decompose the heat current autocorrelator in a similar way:
\begin{equation}
K^\tau_{QQ}(\tau>0,T)=-\sum\limits_{\mathclap{i,j=1}}^5\big\langle\hat{I}_i(\tau)\hat{I}_j(0)\big\rangle\equiv\sum\limits_{\mathclap{i,j=1}}^5C_{ij}(\tau).
\end{equation}
The main task is thus the identification and subsequent evaluation of all non-zero components of $C_{ij}(\tau)$, most of which are three-loop diagrams. Using Wick's theorem, we find that all terms except the diagonal components $C_{ii}$ and the combination $(C_{24}+C_{42})$ vanish due to the fact that they are proportional to bubble diagrams; the interested reader can verify this explicitly with the methods that are also used below. We will now discuss each of the remaining components separately.
\begin{itemize}[wide=0pt,listparindent=\parindent,parsep=0pt]
    \item \textbf{Diagonal component $\pmb{C_{11}}$}\\
    Using Wick's theorem together with the fact that the $\nu=c,s$ modes are decoupled from the $\nu=sf$ modes and the dot, the first component can be written as
    \begin{align}
    C_{11}(\tau)&=\frac{(\pi v_Fg_\bot)^2}{4L^3}\sum\limits_{\mathclap{\substack{k,k^\prime,k^{\prime\prime}\\q,q^\prime,q^{\prime\prime}}}}\left\langle\big(\psi^\dagger_{sf,k}(\tau)-\psi_{sf,k}(\tau)\big)\big(d^\dagger(\tau)-d(\tau)\big)\big(\psi^\dagger_{sf,q}(0)-\psi_{sf,q}(0)\big)\big(d^\dagger(0)-d(0)\big)\right\rangle\nonumber\\
    &\quad\,\times\left\langle\big(\psi_{c,k^\prime}^\dagger(\tau)\psi_{c,k^{\prime\prime}}(\tau)+\psi_{s,k^\prime}^\dagger(\tau)\psi_{s,k^{\prime\prime}}(\tau)\big)\big(\psi_{c,q^\prime}^\dagger(0)\psi_{c,q^{\prime\prime}}(0)+\psi_{s,q^\prime}^\dagger(0)\psi_{s,q^{\prime\prime}}(0)\big)\right\rangle.
    \end{align}
    To simplify the second line, we refer to the previous statement that bubble diagrams vanish, such that the excitation densities corresponding to the $\nu=c,s$ modes are equal to zero. The cross terms do therefore not contribute. Carefully applying Wick's theorem and the definitions of the Green functions, we find
    \begin{align}
    C_{11}(\tau)&=-\frac{(\pi v_Fg_\bot)^2}{4L^3}\sum\limits_{\substack{k,k^\prime,k^{\prime\prime}\\q,q^\prime,q^{\prime\prime}}}\sum_{\mu\nu}^\prime\sum_{\rho\sigma}^\prime\big(G_{ld,k,\mu\nu}(\tau)G_{ld,q,\rho\sigma}(-\tau)+G_{ll,kq,\mu\nu}(\tau)G_{dd,\rho\sigma}(\tau)\big)\nonumber\\
    &\quad\,\times\big(G_{cc,k^\prime q^{\prime\prime},22}(\tau)G_{cc,k^{\prime\prime}q^\prime,11}(\tau)+G_{ss,k^\prime q^{\prime\prime},22}(\tau)G_{ss,k^{\prime\prime}q^\prime,11}(\tau)\big).
    \end{align}
    From Eq.~(\ref{eq:signedsums2}) it follows that the first term on the right-hand side is odd in both $k$ and $q$, and therefore vanishes upon integrating over these momenta. Transformed to bosonic Matsubara frequencies $\Omega_n$, the above thus becomes
    \begin{align}
    C_{11}(i\Omega_n)&=-\frac{(\pi v_Fg_\bot)^2}{4L^3}\frac{1}{(\hbar\beta)^3}\sum\limits_{\substack{k,k^\prime,k^{\prime\prime}\\q,q^\prime,q^{\prime\prime}}}\sum_{\mu\nu}^\prime\sum_{\rho\sigma}^\prime\sum\limits_{n^\prime,n^{\prime\prime},n^{\prime\prime\prime}}G_{ll,kq,\mu\nu}\big(-i(\omega_{n^\prime}+\omega_{n^{\prime\prime}}+\omega_{n^{\prime\prime\prime}}-\Omega_n)\big)G_{dd,\rho\sigma}(i\omega_{n^{\prime\prime\prime}})\nonumber\\
    &\quad\,\times\Big(G_{cc,k^\prime q^{\prime\prime},22}(i\omega_{n^\prime})G_{cc,k^{\prime\prime}q^\prime,11}(i\omega_{n^{\prime\prime}})+G_{ss,k^\prime q^{\prime\prime},22}(i\omega_{n^\prime})G_{ss,k^{\prime\prime}q^\prime,11}(i\omega_{n^{\prime\prime}})\Big),
    \end{align}
    where the sums over $n^\prime,n^{\prime\prime}$ and $n^{\prime\prime\prime}$ all go from $-\infty$ to $\infty$. Since the $\nu=c,s$ modes are completely decoupled, the corresponding Green functions satisfy $\mathbf{G}_{cc,kk^\prime}(i\omega_n)=\mathbf{G}_{ss,kk^\prime}(i\omega_n)=\delta_{k,k^\prime}\mathbf{L}_{0,k}(i\omega_n)$, see Eq.~(\ref{eq:L0k}). Plugging in the expressions from Eqs.~(\ref{eq:signedsums1}) and (\ref{eq:signedsums3}), omitting the terms that are odd in any of the momenta and relabeling the remaining momenta:
    \begin{align}
    C_{11}(i\Omega_n)&=\frac{2(\pi v_Fg_\bot)^2}{(L\beta)^3}\sum\limits_{k,k^\prime,k^{\prime\prime}}\sum\limits_{n^\prime,n^{\prime\prime},n^{\prime\prime\prime}}\frac{1}{i\hbar\omega_{n^\prime}-\epsilon_k}\frac{1}{i\hbar\omega_{n^{\prime\prime}}-\epsilon_{k^\prime}}\frac{1}{i\hbar(\omega_{n^\prime}+\omega_{n^{\prime\prime}}+\omega_{n^{\prime\prime\prime}}-\Omega_n)-\epsilon_{k^{\prime\prime}}}D_{bb}(i\omega_{n^{\prime\prime\prime}}).
    \end{align}
    
    Having found an explicit formula for the three-loop diagram $C_{11}(i\Omega_n)$, we continue by evaluating two of the Matsubara sums. We do so by using the following identity for the Fermi-Dirac distribution $n_F(\epsilon)$:
    \begin{equation}
    \frac{1}{\beta}\sum\limits_{\mathclap{n=-\infty}}^\infty\frac{1}{i\hbar\omega_n-\epsilon}\frac{1}{i\hbar\omega_n-\epsilon^\prime}=\frac{n_F(\epsilon)-n_F(\epsilon^\prime)}{\epsilon-\epsilon^\prime}.\label{eq:matsum}
    \end{equation}
    Furthermore, it is straightforward to show that $n_F(\epsilon-i\hbar\Omega_n)=n_F(\epsilon)$ and $n_F(\epsilon-i\hbar\omega_n)=-n_B(\epsilon)$ for bosonic and fermionic Matsubara frequencies, respectively, where $n_B(\epsilon)$ is the Bose-Einstein distribution. Applying Eq.~(\ref{eq:matsum}) twice and taking the continuum limit of all momentum sums, we obtain
    \begin{align}
    C_{11}(i\Omega_n)&=\frac{\Gamma}{8\pi\hbar^2\beta}\int\limits_{-\infty}^\infty\mathrm{d}\epsilon_k\int\limits_{-\infty}^\infty\mathrm{d}\epsilon_{k^\prime}\int\limits_{-\infty}^\infty\mathrm{d}\epsilon_{k^{\prime\prime}}\sum\limits_{\mathclap{n^\prime=-\infty}}^\infty D_{bb}(i\omega_{n^\prime})\frac{\big(n_F(\epsilon_{k^\prime})-n_F(\epsilon_{k^{\prime\prime}})\big)\big(n_F(\epsilon_k)+n_B(\epsilon_{k^{\prime\prime}}-\epsilon_{k^\prime})\big)}{i\hbar\omega_{n^\prime-n}-(\epsilon_{k^{\prime\prime}}-\epsilon_k-\epsilon_{k^\prime})}.
    \end{align}
    Also switching to new variables $\epsilon\equiv(\epsilon_k+\epsilon_{k^\prime}-\epsilon_{k^{\prime\prime}})/2$, $\epsilon^\prime\equiv(\epsilon_k-\epsilon_{k^\prime}-\epsilon_{k^{\prime\prime}})/2$, $\epsilon^{\prime\prime}\equiv\epsilon_k+\epsilon_{k^\prime}+\epsilon_{k^{\prime\prime}}$:
    \begin{align}
    C_{11}(i\Omega_n)&=\frac{\Gamma}{8\pi\hbar^2\beta}\int\limits_{-\infty}^\infty\mathrm{d}\epsilon\int\limits_{-\infty}^\infty\mathrm{d}\epsilon^\prime\int\limits_{-\infty}^\infty\mathrm{d}\epsilon^{\prime\prime}\sum\limits_{\mathclap{n^\prime=-\infty}}^\infty D_{bb}(i\omega_{n^\prime})\nonumber\\
    &\quad\,\times\frac{\big(n_F(\epsilon-\epsilon^\prime)-n_F(-\epsilon+\epsilon^{\prime\prime}/2)\big)\big(n_F(\epsilon^\prime+\epsilon^{\prime\prime}/2)+n_B(-2\epsilon+\epsilon^\prime+\epsilon^{\prime\prime}/2)\big)}{i\hbar\omega_{n^\prime-n}+2\epsilon}\nonumber\\
    &=\frac{\Gamma}{4\pi\hbar^2\beta}\int\limits_{-\infty}^\infty\mathrm{d}\epsilon\int\limits_{-\infty}^\infty\mathrm{d}\epsilon^\prime\sum\limits_{\mathclap{n^\prime=-\infty}}^\infty\frac{(\epsilon+\epsilon^\prime)\cosh(\beta\epsilon)}{\sinh(\beta\epsilon)+\sinh(\beta\epsilon^\prime)}\frac{1}{i\hbar\omega_{n^\prime-n}+2\epsilon}D_{bb}(i\omega_{n^\prime})\nonumber\\
    &=\frac{\Gamma}{4\pi\hbar^2\beta}\int\limits_{-\infty}^\infty\mathrm{d}\epsilon\sum\limits_{\mathclap{n^\prime=-\infty}}^\infty\left(\frac{\pi^2}{2\beta^2}+2\epsilon^2\right)\frac{1}{i\hbar\omega_{n^\prime-n}+2\epsilon}D_{bb}(i\omega_{n^\prime})\nonumber\\
    &\rightarrow-\frac{\Gamma}{4\pi\hbar\beta}\int\limits_{-\Lambda^\prime}^{\Lambda^\prime}\mathrm{d}\epsilon\sum\limits_{\mathclap{n^\prime=-\infty}}^\infty\left(\frac{\pi^2}{2\beta^2}+2\epsilon^2\right)\frac{\hbar\omega_{n^\prime-n}}{(\hbar\omega_{n^\prime-n})^2+(2\epsilon)^2}\frac{1}{\hbar\omega_{n^\prime}+\text{sgn}(\omega_{n^\prime})\Gamma},\label{eq:C11ex}
    \end{align}
    where $\Lambda^\prime=3\Lambda/2$ is the cut-off of the redefined variable $\epsilon$, and we used Eq.~(\ref{eq:Dbbomega}) for the dot Green function. Next, we write out the Matsubara frequencies explicitly, perform the final integral, and take the limit $\Lambda^\prime\rightarrow\infty$ to find
    \begin{equation}
    C_{11}(i\Omega_n)=-\frac{\Gamma}{16\pi\hbar\beta^2}\sum\limits_{\mathclap{n^\prime=-\infty}}^\infty\frac{\pi^2\text{sgn}\!\left(n^\prime-n+\frac{1}{2}\right)\left(\frac{1}{2}-2\left(n^\prime-n+\frac{1}{2}\right)^2\right)+4\beta\Lambda^\prime\left(n^\prime-n+\frac{1}{2}\right)}{n^\prime+\frac{1}{2}+\text{sgn}\!\left(n^\prime+\frac{1}{2}\right)\frac{\beta\Gamma}{2\pi}}.
    \end{equation}
    Note that this result for the integral assumes that $\omega_{n^\prime}$ remains finite, which is not true for all terms of the sum. The actual expression involves objects such as $\arctan(\Lambda^\prime/\hbar\omega_{n^\prime-n})$, effectively introducing a cut-off $N$ in the sum over $n^\prime$. Although the naive introduction of a hard cut-off $N$ does lead to errors in the expression for the current autocorrelator $K^\tau_{QQ}(i\Omega_{n>0},T)$, the desired dc limit of the linear susceptibility is still exact due to the fact that the erroneous region $\hbar\omega_{n^\prime}\sim \Lambda^\prime$ does not contribute to the linear order term in $n$. The latter follows from the fact that the autocorrelator can be rewritten to only contain the combination $D_{bb}(i\omega_{n^\prime-n})-D_{bb}(-i\omega_{n^\prime+n})$: for terms in the region $\hbar\omega_{n^\prime}\sim \Lambda^\prime\rightarrow\infty$ (i.e. $n^\prime\gg n$), this combination is both analytic and even in $n$, see Eq.~(\ref{eq:Dbbomega}). The errors introduced by writing $\arctan(\Lambda^\prime/\hbar\omega_{n^\prime-n})\rightarrow \text{sgn}(\omega_{n^\prime-n})\,\pi/2$ therefore only depend on even powers of $n$.
    
    The most obvious way to calculate the linear susceptibility is to expand the current autocorrelator in $n$ and extract the linear part. However, this is only possible if the correlator is analytic, which the summand of the above expression is not. To work around this, we split the sum into different parts in which the sign functions reduce to constants. Restricting ourselves to $n>0$, the three different parts are: (i) $n^\prime<0$, with both sign functions equal to $-1$; (ii) $0\leq n^\prime<n$, where one of the sign functions is $-1$ while the other is $+1$; (iii) $n^\prime\geq n$, with both sign functions equal to $+1$. Writing $n^\prime\rightarrow-n^\prime-1$ in the first part, using $\sum_{n^\prime=n}^\infty=\sum_{n^\prime=0}^\infty-\sum_{n^\prime=0}^{n-1}$ in the third part, and subsequently combining the parts that sum over $n^\prime\in\{0,\ldots,n-1\}$, we obtain the following analytic form:
    \begin{align}
    C_{11}(i\Omega_{n>0})&=-\frac{\Gamma}{16\pi\hbar\beta^2}\Bigg(-2\pi^2\sum\limits_{n^\prime=0}^{n-1}\frac{\frac{1}{2}-2\left(n^\prime-n+\frac{1}{2}\right)^2}{n^\prime+\frac{1}{2}+\frac{\beta\Gamma}{2\pi}}+\sum\limits_{n^\prime=0}^\infty\frac{\pi^2\left(\frac{1}{2}-2\left(n^\prime+n+\frac{1}{2}\right)^2\right)+4\beta\Lambda^\prime\left(n^\prime+n+\frac{1}{2}\right)}{n^\prime+\frac{1}{2}+\frac{\beta\Gamma}{2\pi}}\nonumber\\
    &\quad\,+\sum\limits_{n^\prime=0}^\infty\frac{\pi^2\left(\frac{1}{2}-2\left(n^\prime-n+\frac{1}{2}\right)^2\right)+4\beta\Lambda^\prime\left(n^\prime-n+\frac{1}{2}\right)}{n^\prime+\frac{1}{2}+\frac{\beta\Gamma}{2\pi}}\Bigg).
    \end{align}
    The second and third sums of this expression diverge, being proportional to $\Lambda^2$. However, these lines combined only contain terms that are either constant or quadratic in $n$. For the purpose of finding the linear susceptibility, the above autocorrelator therefore simplifies to
    \begin{equation}
     C_{11}(i\Omega_{n>0})=\text{const.}+\frac{\pi\Gamma}{8\hbar\beta^2}\sum\limits_{n^\prime=0}^{n-1}\frac{\frac{1}{2}-2\left(n^\prime-n+\frac{1}{2}\right)^2}{n^\prime+\frac{1}{2}+\frac{\beta\Gamma}{2\pi}}+\mathcal{O}(\Omega_n^2).
    \end{equation}
    Finally evaluating the remaining sum, expanding the result to linear order in $n$, and performing analytic continuation to real frequencies, we find
    \begin{equation}
    C^\text{R}_{11}(\omega)=\text{const.}-\frac{i\Gamma}{16\hbar\beta}\left[\frac{\beta\Gamma}{\pi}+\left(\frac{1}{2}-\frac{\beta^2\Gamma^2}{2\pi^2}\right)\psi^{(1)}\left(\frac{1}{2}+\frac{\beta\Gamma}{2\pi}\right)\right]\hbar\omega+\mathcal{O}(\omega^2).
    \end{equation}
    If we furthermore identify $\beta\Gamma$ as $T_K/T\rightarrow\infty$ and utilize the expansion of the trigamma function
    \begin{equation}
    \frac{1}{x}\psi^{(1)}\left(\frac{1}{2}+\frac{1}{x}\right)=1-\frac{x^2}{12}+\mathcal{O}\left(x^4\right),\label{eq:trigammaexp}
    \end{equation}
    we find that this term of the heat current autocorrelator reduces to
    \begin{equation}
    C^\text{R}_{11}(\omega)=\text{const.}-\frac{i\pi\omega}{12\beta^2}+\mathcal{O}(\omega^2)\label{eq:C11Tsquare}
    \end{equation}
    at the NFL fixed point.
    \item \textbf{The $\pmb{\nu=f}$ terms: $\pmb{C_{22}+C_{44}+C_{24}+C_{42}}$}\\
    Going through the same procedure as for $C_{11}$ and using that the sum over all components of $\mathbf{G}_{ld,k}(i\omega_n)$ is equal to zero, we find
    \begin{align}
    C_{22}(i\Omega_n)&=-\frac{(\pi v_Fg_\bot)^2}{4L^3}\frac{1}{(\hbar\beta)^3}\sum\limits_{\substack{k,k^\prime,k^{\prime\prime}\\q,q^\prime,q^{\prime\prime}}}\sum_{\mu\nu}\sum_{\rho\sigma}\sum\limits_{n^\prime,n^{\prime\prime},n^{\prime\prime\prime}}G_{ff,k^\prime q^{\prime\prime},22}(i\omega_{n^\prime})G_{ff,k^{\prime\prime}q^\prime,11}(i\omega_{n^{\prime\prime}})\nonumber\\
    &\quad\,\times G_{ll,kq,\mu\nu}(i\omega_{n^{\prime\prime\prime}})G_{dd,\rho\sigma}\big(-i(\omega_{n^\prime}+\omega_{n^{\prime\prime}}+\omega_{n^{\prime\prime\prime}}-\Omega_n)\big)\nonumber
    \end{align}
    \begin{align}
    &=\frac{\hbar(\pi v_Fg_\bot)^2}{(L\beta)^3}\sum\limits_{k,k^\prime,k^{\prime\prime}}\sum\limits_{n^\prime,n^{\prime\prime},n^{\prime\prime\prime}}\left(1+\frac{4g_\bot^2}{\hbar L}\sum\limits_{k^{\prime\prime\prime}}\frac{1}{i\hbar\omega_{n^{\prime\prime\prime}}-\epsilon_{k^{\prime\prime\prime}}}D_{bb}(i\omega_{n^{\prime\prime\prime}})\right)\nonumber\\
    &\quad\,\times\frac{1}{i\hbar\omega_{n^\prime}-\epsilon_k}\frac{1}{i\hbar\omega_{n^{\prime\prime}}-\epsilon_{k^\prime}}\frac{1}{i\hbar\omega_{n^{\prime\prime\prime}}-\epsilon_{k^{\prime\prime}}}\frac{1}{i\hbar(\omega_{n^\prime}+\omega_{n^{\prime\prime}}+\omega_{n^{\prime\prime\prime}}-\Omega_n)}.
    \end{align}
    Also evaluating the sums in the same way as for $C_{11}$ (i.e. performing two frequency sums using Eq.~(\ref{eq:matsum}), taking the continuum limit of the momentum sums, introducing the coordinates $\epsilon\equiv(\epsilon_k+\epsilon_{k^\prime})/2$, $\epsilon^\prime\equiv\epsilon_k-\epsilon_{k^\prime}$, and evaluating the integrals over $\epsilon_{k^{\prime\prime}}$, $\epsilon_{k^{\prime\prime\prime}}$ and $\epsilon^\prime$):
    \begin{equation}
    C_{22}(i\Omega_n)=-\frac{\Gamma}{8\hbar\beta}\int\limits_{-\Lambda}^{\Lambda}\mathrm{d}\epsilon\sum\limits_{\mathclap{n^\prime=-\infty}}^\infty\frac{\epsilon}{\tanh(\beta\epsilon)}\frac{\hbar\omega_{n^\prime-n}}{(\hbar\omega_{n^\prime-n})^2+(2\epsilon)^2}\frac{\hbar\omega_{n^\prime}}{|\hbar\omega_{n^\prime}|+\Gamma}.\label{eq:C22}
    \end{equation}
    Before going any further, we also calculate the component
    \begin{equation}
    C_{44}(\tau)=\frac{(\pi v_F)^2}{4L^2}\sum\limits_{\mathclap{\substack{k,k^\prime\\q,q^\prime}}}(\epsilon_{k^\prime}-\epsilon_k)(\epsilon_{q^\prime}-\epsilon_q)\big\langle a(\tau)a(0)\big\rangle\big\langle b(\tau)b(0)\big\rangle\left\langle\psi_{f,k}^\dagger(\tau)\psi_{f,k^\prime}(\tau)\psi_{f,q}^\dagger(0)\psi_{f,q^\prime}(0)\right\rangle.
    \end{equation}
    Once again following the same procedure as for the previous components, this becomes
    \begin{align}
    C_{44}(i\Omega_n)&=\frac{(\pi v_F)^2}{4L^2\beta^3}\sum\limits_{\mathclap{k,k^\prime}}\sum\limits_{n^\prime,n^{\prime\prime},n^{\prime\prime\prime}}(\epsilon_{k^\prime}-\epsilon_k)^2\frac{1}{i\hbar\omega_{n^\prime}+\epsilon_k}\frac{1}{i\hbar\omega_{n^{\prime\prime}}-\epsilon_{k^\prime}}\frac{1}{i\hbar(\omega_{n^\prime}+\omega_{n^{\prime\prime}}+\omega_{n^{\prime\prime\prime}}-\Omega_n)}D_{bb}(i\omega_{n^{\prime\prime\prime}})\nonumber\\
    &=-\frac{1}{2\hbar\beta}\int\limits_{-\Lambda}^{\Lambda}\mathrm{d}\epsilon\sum\limits_{\mathclap{n^\prime=-\infty}}^\infty\frac{\epsilon^3}{\tanh(\beta\epsilon)}\frac{\hbar\omega_{n^\prime-n}}{(\hbar\omega_{n^\prime-n})^2+(2\epsilon)^2}\frac{1}{\hbar\omega_{n^\prime}+\text{sgn}(\omega_{n^\prime})\Gamma}.\label{eq:C44int}
    \end{align}
    Finally, without explicitly going through the calculation, the combination $(C_{24}+C_{42})$ can analogously be derived to be equal to
    \begin{equation}
    C_{24}(i\Omega_n)+C_{42}(i\Omega_n)=-\frac{\Gamma}{\hbar\beta}\int\limits_{-\Lambda}^{\Lambda}\mathrm{d}\epsilon\sum\limits_{\mathclap{n^\prime=-\infty}}^\infty\frac{\epsilon^3}{\tanh(\beta\epsilon)}\frac{1}{(\hbar\omega_{n^\prime-n})^2+(2\epsilon)^2}\frac{1}{|\hbar\omega_{n^\prime}|+\Gamma}.\label{eq:C2442}
    \end{equation}
    
    We now extract the contribution of the above four components to the linear susceptibility by combining the components and discussing them together, starting with Eqs.~(\ref{eq:C22}) and (\ref{eq:C2442}). Combined, these first three terms can be written as
    \begin{align}
    C_{22}(i\Omega_n)+C_{24}(i\Omega_n)+C_{42}(i\Omega_n)&=-\frac{\Gamma}{2\hbar\beta}\int\limits_{-\Lambda}^{\Lambda}\mathrm{d}\epsilon\sum\limits_{\mathclap{n^\prime=-\infty}}^\infty\frac{\epsilon^3}{\tanh(\beta\epsilon)}\frac{1}{(\hbar\omega_{n^\prime-n})^2+(2\epsilon)^2}\frac{1}{|\hbar\omega_{n^\prime}|+\Gamma}\nonumber\\
    &\quad\,-\frac{\Gamma}{8\hbar\beta}\int\limits_{-\Lambda}^{\Lambda}\mathrm{d}\epsilon\sum\limits_{\mathclap{n^\prime=-\infty}}^\infty\frac{\epsilon}{\tanh(\beta\epsilon)}\frac{1}{|\hbar\omega_{n^\prime}|+\Gamma}\nonumber\\
    &\quad\,-\frac{\Gamma\Omega_n}{8\beta}\int\limits_{-\Lambda}^{\Lambda}\mathrm{d}\epsilon\sum\limits_{\mathclap{n^\prime=-\infty}}^\infty\frac{\epsilon}{\tanh(\beta\epsilon)}\frac{\hbar\omega_{n^\prime-n}}{(\hbar\omega_{n^\prime-n})^2+(2\epsilon)^2}\frac{1}{|\hbar\omega_{n^\prime}|+\Gamma}.\label{eq:C222442}
    \end{align}
    The final two lines of this expression do not contribute to the linear susceptibility: the second term does not depend on $n$ at all, while the third term is at least quadratic on $\Omega_n$ (to see this, simply note that the summand is odd in $\omega_{n^\prime}$ if $n=0$). With that in mind, we unite the four components. Splitting the remaining sums over $n^\prime$ into an $n^\prime<0$ part and an $n^\prime\geq 0$ part, and writing $n^\prime\rightarrow-n^\prime-1$ in the former, we find
    \begin{align}
    C_{22}(i\Omega_n)+C_{44}(i\Omega_n)+C_{24}(i\Omega_n)+C_{42}(i\Omega_n)&=\text{const.}-\frac{1}{2\hbar\beta}\int\limits_{-\Lambda}^{\Lambda}\mathrm{d}\epsilon\sum\limits_{n^\prime=0}^\infty\frac{\epsilon^3}{\tanh(\beta\epsilon)}\frac{1}{\hbar\omega_{n^\prime}+\Gamma}\nonumber\\
    &\quad\,\times\left(\frac{\hbar\omega_{n^\prime+n}+\Gamma}{(\hbar\omega_{n^\prime+n})^2+(2\epsilon)^2}+\frac{\hbar\omega_{n^\prime-n}+\Gamma}{(\hbar\omega_{n^\prime-n})^2+(2\epsilon)^2}\right)+\mathcal{O}(\Omega_n^2).\label{eq:Ccombi}
    \end{align}
    Contrary to the previously calculated autocorrelators, the remaining integral cannot be evaluated exactly. As such, we are required to expand in $n$ before having evaluated all of the sums and integrals. Formally, this is the incorrect order of operations, therefore leading to incorrect results if not done carefully. For example, although Eq.~(\ref{eq:Ccombi}) seems to imply that the remaining sum only contributes to even powers of $n$, this is not necessarily true. The reason for this is hidden in the fact that $\omega_{n^\prime-n}<0$ for some of the terms, such that the usual identities involving digamma and trigamma functions cannot be applied directly. As a result, the sum over the terms involving $\omega_{n^\prime-n}$ evaluates to a different analytic function than the sum over the terms that depend on $\omega_{n^\prime+n}$. The evaluated sum is thus of the form $(f(-n)+g(n))$ instead of $(f(-n)+f(n))$, therefore generally supplying odd powers of $n$ as well. Taking this into account, we have to explicitly evaluate the sum before expanding it in $n$. Doing so, we find that the resulting power series does indeed contain odd powers of $n$, but the linear term is missing. The combination of components from Eq.~(\ref{eq:Ccombi}) does therefore not contribute to the linear susceptibility.
    \item \textbf{Diagonal component $\pmb{C_{33}}$}\\
    The component $C_{33}$ is very similar to $C_{22}$, such that we can straightforwardly modify the previous steps to find
    \begin{align}
    C_{33}(i\Omega_n)&=-\frac{(\Lambda g_\bot)^2}{16\hbar^2L}\frac{1}{\hbar\beta}\sum\limits_{\mathclap{k,k^\prime}}\sum_{\mu\nu}\sum_{\rho\sigma}\sum\limits_{n^\prime=-\infty}^\infty G_{ll,kk^\prime,\mu\nu}(i\omega_{n^\prime})G_{dd,\rho\sigma}(-i\omega_{n^\prime-n})\nonumber\\
    &=-\frac{\Gamma\Lambda^2}{16\hbar\beta}\sum\limits_{\mathclap{n^\prime=-\infty}}^\infty\frac{1}{\hbar\omega_{n^\prime-n}}\frac{\hbar\omega_{n^\prime}}{|\hbar\omega_{n^\prime}|+\Gamma}.
    \end{align}
    Evaluating the sum in the same fashion as before, it can be shown that the latter sum does not contain a linear term in $n$. Consequently, this component does also not contribute to the linear susceptibility.
    \item \textbf{Diagonal component $\pmb{C_{55}}$}\\
    This component is by far the most complicated due to the fact that the $\nu=sf$ modes are coupled to the $b$ Majorana mode, combined with the fact that the propagators corresponding to these modes contain non-zero off-diagonal components. Keeping that in mind, Wick's theorem gives us 15 terms to consider. Five of these terms are vanishing bubble diagrams, while the remaining four bubble diagrams do not have a linear term. For the purpose of finding the linear susceptibility, we therefore only have to consider six terms. Without explicitly performing the lengthy calculation, we note that these combined terms can be expressed in the following way:
    \begin{align}
    C_{55}(i\Omega_n)&=\text{const.}+C_{44}(i\Omega_n)-\frac{(\pi\hbar v_Fg_\bot)^2}{(L\hbar\beta)^3}\sum\limits_{k,k^\prime,k^{\prime\prime}}\sum\limits_{n,n^\prime,n^{\prime\prime}}(\epsilon_k-\epsilon_{k^\prime})(\epsilon_k-\epsilon_{k^{\prime\prime}})\frac{1}{i\hbar\omega_{n^{\prime\prime\prime}}-\epsilon_k}\frac{1}{i\hbar\omega_{n^{\prime\prime}}+\epsilon_{k^{\prime\prime}}}\nonumber\\
    &\quad\,\times\left(\frac{1}{i\hbar\omega_{n^\prime}+\epsilon_{k^\prime}}-\frac{1}{i\hbar\omega_{n^{\prime\prime}}+\epsilon_{k^\prime}}\right)\frac{1}{i\hbar(\omega_{n^\prime}+\omega_{n^{\prime\prime}}+\omega_{n^{\prime\prime\prime}}-\Omega_n)}D_{bb}(i\omega_{n^\prime})D_{bb}(i\omega_{n^{\prime\prime}})+\mathcal{O}(\Omega_n^2).
    \end{align}
    It can be shown that the isolated component $C_{44}$ does in fact contain a linear term in $\Omega_n$, however, this term goes to zero with $T/T_K$. As such, $C_{44}$ does not contribute to the linear susceptibility at this point, and we can instead focus on the other terms.
    \begin{figure}[t]
    \centerline{\begin{tabular}{ccc}
    \includegraphics[width=5.8cm]{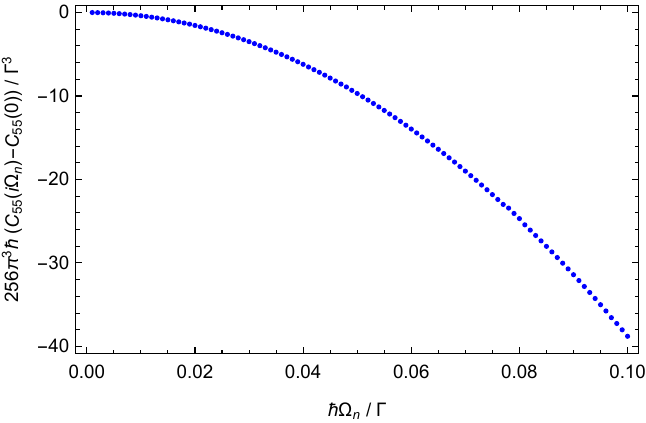} & \hspace{1cm} & \includegraphics[width=5.8cm]{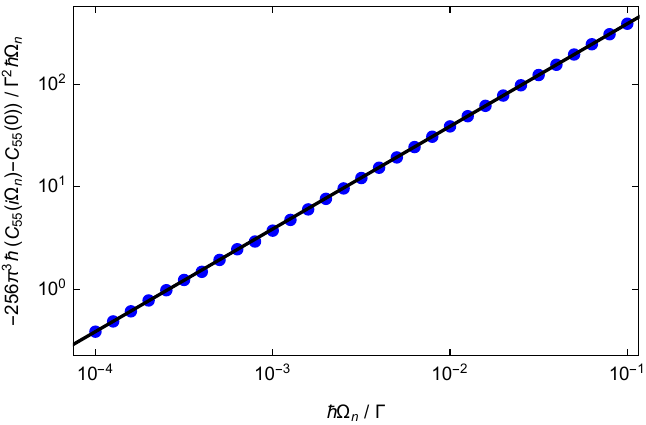}
    \end{tabular}}
    \caption{\label{fig:C55}The component $C_{55}(i\Omega_n)$ at the NFL fixed point, numerically calculated as a function of dimensionless Matsubara frequency $\hbar\Omega_n/\Gamma$ with $\Lambda/\Gamma=10^2$. Left: $C_{55}(i\Omega_n)$ minus its zeroth order term, rescaled with a constant prefactor to make it dimensionless. Right: log-log plot of minus the same object, divided by the dimensionless frequency. The solid line is a function of the form $y=ax$ (its slope in the log-log plot therefore being equal to $1$), confirming that the susceptibility is perfectly linear in the frequency over this domain. Note that these curves are independent of temperature in the regime $T\ll T_K$.}
    \end{figure}

    Contrary to all of the previously calculated terms, the remaining terms cannot be calculated exactly, nor can they be successfully expanded in $\Omega_n$ before evaluation. The reason for this is the presence of an additional $D_{bb}$ propagator that is interwoven in the sums. Instead of using analytical methods, we therefore calculate the sums numerically as a function of $\Omega_n$, and show that the corresponding contribution to the linear susceptibility goes to zero at the NFL fixed point. The results for $\beta\Gamma\rightarrow\infty$ (i.e. at the NFL fixed point) are shown in Fig.~\ref{fig:C55}, where we have set the only remaining parameter $\Lambda/\Gamma$ to $10^2$ as an example. As can be deduced from the left panel, the lowest non-trivial order term of the component $C_{55}(i\Omega_n)$ is quadratic in $\Omega_n$, similar to what we have seen for most of the other components. In addition, the right panel shows a log-log plot of the corresponding contribution to the linear susceptibility $\chi_{55}(i\Omega_n)$ up to a constant prefactor. Upon analytically continuing the data to real frequencies, the plot confirms that this contribution to the susceptibility is perfectly linear in $\omega$ over the entire small-$\omega$ region, such that it goes to zero in the dc limit $\omega\rightarrow 0$.
\end{itemize}
To summarize, we have shown that only the component $C_{11}$ has a linear term in the frequency at the NFL fixed point. Explicitly, we thus find that the full NFL heat current autocorrelator is given by
\begin{equation}
K_{QQ}(\omega,T)=\text{const.}-\frac{i\pi\omega}{12\beta^2}+\mathcal{O}(\omega^2).
\end{equation}
From Eq.~(\ref{eq:kubo}), we now finally obtain the following dc heat susceptibility:
\begin{equation}
\chi_{QQ}=\frac{\pi^2k_B^2T^2}{6h}.
\end{equation}

We also briefly comment on the off-diagonal terms $\chi_{Qc}$ and $\chi_{cQ}$. Referring back to Eqs.~(\ref{eq:Ic}) and (\ref{eq:I1})-(\ref{eq:I5}), we immediately see that any terms involving $\hat{I}_1$, $\hat{I}_2$ or $\hat{I}_4$ are proportional to vanishing bubble diagrams. Moreover, the charge current operator does not contain the $a$ Majorana fermion, such that the products of $\hat{I}_c$ with either $\hat{I}_3$ or $\hat{I}_5$ contain exactly one $a$ operator. At the NFL fixed point, the $a$ Majorana fermion is completely decoupled from all other modes, and all terms involving $\hat{I}_3$ and $\hat{I}_5$ are therefore equal to zero as well. We thus conclude that the off-diagonal terms $\chi_{Qc}$ and $\chi_{cQ}$ are equal to zero at the NFL fixed point, and as such the temperature gradient does not induce thermopower. Consequently, the two choices $V=0$ and $I_c=0$ coincide, such that the heat conductance $\kappa$ is unambiguously given by
\begin{equation}
\kappa=\frac{\chi_{QQ}}{T}=\frac{\pi^2k_B^2T}{6h}\label{eq:heatc}
\end{equation}
at the NFL fixed point of the C2CK model. This is the main result from Eq.~(\ref{eq:main}).

\section*{Corrections to the Emery-Kivelson point charge conductance}
\begin{fmffile}{diagrams}
We explicitly calculate the corrections to the linear response charge conductance away from the EK point to lowest order in $\lambda\equiv 2\pi\hbar v_F-J_z$ and $T/T_K$. Our starting point is the interaction term from Ref.~\cite{emery1992mapping},
\begin{align}
\hat{H}_I&=\lambda:\psi_s^\dagger(0)\psi_s(0):\left(d^\dagger d-\frac{1}{2}\right)\nonumber\\
&=\frac{i\lambda}{L}ba\sum_{k,k^\prime}:\psi_{s,k}^\dagger\psi_{s,k^\prime}:,\label{eq:Hint}
\end{align}
which we will treat as a perturbation to the non-interacting Hamiltonian from Eq.~(\ref{eq:HEK})~\cite{majumdar1998perturbation}. First, we consider the NFL (i.e. leading order) charge current autocorrelator; repeating a much simpler version of the calculations for the heat conductance shown above, we find that it is given by
\begin{equation}
K^\tau_{cc}(i\Omega_{n>0},T)=-\frac{e^2\Gamma}{8\pi\hbar^3\beta}\sum\limits_{n^\prime=-\infty}^\infty\int\limits_{-\infty}^\infty\mathrm{d}\epsilon_k\text{Tr}\!\left[\mathbf{L}_{0,k}(i\omega_{n^\prime})\right]D_{bb}(-i\omega_{n^\prime-n}).\label{eq:corrD}
\end{equation}
Since the interaction term does not involve $\nu=sf$ modes, the bare propagators corresponding to those modes remain unchanged. Our first objective is thus to find the corrections to the $bb$ component of the dot Green function in presence of a non-zero $\lambda$.

We approximate the full $bb$ component of the dot Green function $D_{bb}^\text{full}(i\omega_n)$ in presence of interactions by employing standard Feynman techniques. Using that Eq.~(\ref{eq:Hint}) provides a four-point vertex involving two $\psi_{s,k}$ legs, an $a$ leg and a $b$ leg, the Feynman rules lead to the following diagrammatic expression for $D_{bb}^\text{full}(i\omega_n)$:
\begin{equation}
\parbox{20mm}{\begin{fmfgraph*}(50,60) \fmfleft{i} \fmfright{o} \fmf{double,label=$\omega_n$}{i,o}\end{fmfgraph*}}=\;\;\;\parbox{20mm}{\begin{fmfgraph*}(50,60) \fmfleft{i} \fmfright{o} \fmf{plain,label=$\omega_n$}{i,o} \end{fmfgraph*}}+\;\;\;\parbox{20mm}{\begin{fmfgraph*}(150,60) \fmfleft{i} \fmfright{o} \fmf{plain,label=$\omega_n$}{i,v1} \fmf{photon,label=$\omega_{n-l+m}$}{v1,v2} \fmf{dashes_arrow,left,label=$\omega_l$,tension=0}{v1,v2} \fmf{dashes_arrow,left,label=$\omega_m$,tension=0}{v2,v1}  \fmf{plain,label=$\omega_n$}{v2,o} \end{fmfgraph*}}\hspace{3.5cm}+\ldots.\label{eq:diagrams}
\end{equation}
Here, each vertex comes with a prefactor $i\lambda/\hbar^2\beta$ and a sum over Matsubara frequencies; the definitions of the other components can be found in Table~\ref{tab:feynman}. Reading off the above Feynman diagrams, we find that the lowest order of the self-energy is given by
\begin{equation}
\Sigma(i\omega_n)=-\frac{\lambda^2}{\hbar^2L^2}\frac{1}{(\hbar\beta)^2}\sum_{n^\prime,n^{\prime\prime}}\sum_{k,k^\prime}D_{aa}\big(-i(\omega_{n^\prime}-\omega_{n^{\prime\prime}}-\omega_n)\big)G_{s,k}(i\omega_{n^\prime})G_{s,k^\prime}(i\omega_{n^{\prime\prime}}),\label{eq:selfenergyfull}
\end{equation}
where $G_{s,k}(i\omega_n)$ is shorthand notation for $G_{ss,kk,11}(i\omega_n)$.
\begin{table}[t]
\centering
\caption{\label{tab:feynman}Definitions of the different components of the Feynman diagrams. The arrow in the fourth diagram indicates the propagation direction of $\psi_{s,k}$.}
\begin{tabular}{|c|c||c|}
     \hline
     {Expression} & {\hspace{0.7cm}Diagram\hspace{0.7cm}} & Vertex \\
     \hline \hline
     {$D_{bb}^\text{full}(i\omega_n)$} & {\begin{fmfgraph*}(50,10) \fmfleft{i} \fmfright{o} \fmf{double,label=$\omega_n$}{i,o} \end{fmfgraph*}} & \\[0.2cm]
     \cline{1-2}
     {$D_{bb}(i\omega_n)$} & {\begin{fmfgraph*}(50,10) \fmfleft{i} \fmfright{o} \fmf{plain,label=$\omega_n$}{i,o} \end{fmfgraph*}} & \multirow{3}{*}{\begin{fmfgraph*}(50,40) \fmfleft{i1,i2} \fmfright{o1,o2} \fmf{dashes_arrow}{i1,v,o2} \fmf{plain}{i2,v} \fmf{photon}{v,o1} \end{fmfgraph*}}\\[0.2cm]
     \cline{1-2}
     {$D_{aa}(i\omega_n)$} & {\begin{fmfgraph*}(50,10) \fmfleft{i} \fmfright{o} \fmf{photon,label=$\omega_n$}{i,o} \end{fmfgraph*}} & \\[0.2cm]
     \cline{1-2}
     {$\frac{1}{L}\sum\limits_k G_{s,k}(i\omega_n)$} & {\begin{fmfgraph*}(50,10) \fmfleft{i} \fmfright{o} \fmf{dashes_arrow,label=$\omega_n$}{i,o} \end{fmfgraph*}} & \\[0.2cm]
     \hline
\end{tabular}
\end{table}

Using that the $a$ and $\psi_{s,k}$ modes are completely isolated from the rest of the system if $\lambda=0$, and taking the continuum limit of the sums over $k$ and $k^\prime$, we have
\begin{align}
\Sigma(i\omega_n)&=-\frac{\lambda^2}{\hbar v_F^2}\frac{1}{\beta^2}\int\limits_{-\infty}^\infty\frac{\mathrm{d}\epsilon_k}{2\pi\hbar}\int\limits_{-\infty}^\infty\frac{\mathrm{d}\epsilon_{k^\prime}}{2\pi\hbar}\sum_{n^\prime,n^{\prime\prime}}\frac{1}{i\hbar(\omega_n-\omega_{n^\prime}+\omega_{n^{\prime\prime}})}\frac{1}{i\hbar\omega_{n^\prime}-\epsilon_k}\frac{1}{i\hbar\omega_{n^{\prime\prime}}-\epsilon_{k^\prime}}.
\end{align}
Furthermore applying Eq.~(\ref{eq:matsum}) twice, together with the substitutions $\epsilon\equiv(\epsilon_k+\epsilon_{k^\prime})/2,\epsilon^\prime\equiv\epsilon_k-\epsilon_{k^\prime}$, the self-energy becomes
\begin{align}
\Sigma(i\omega_n)&=\frac{\lambda^2}{\hbar v_F^2}\int\limits_{-\infty}^\infty\frac{\mathrm{d}\epsilon_k}{2\pi\hbar}\int\limits_{-\infty}^\infty\frac{\mathrm{d}\epsilon_{k^\prime}}{2\pi\hbar}\frac{\big(n_F(0)-n_F(\epsilon_{k^\prime})\big)\big(n_F(\epsilon_k)+n_B(\epsilon_{k^\prime})\big)}{i\hbar\omega_n-(\epsilon_k-\epsilon_{k^\prime})}\nonumber\\
&=\frac{\lambda^2}{4\hbar v_F^2}\int\limits_{-\infty}^\infty\frac{\mathrm{d}\epsilon_k}{2\pi\hbar}\int\limits_{-\infty}^\infty\frac{\mathrm{d}\epsilon_{k^\prime}}{2\pi\hbar}\frac{\cosh\left(\beta(\epsilon_k+\epsilon_{k^\prime})/2\right)}{\cosh\left(\beta\epsilon_k/2\right)\cosh\left(\beta\epsilon_{k^\prime}/2\right)}\frac{1}{i\hbar\omega_n-(\epsilon_k+\epsilon_{k^\prime})}\nonumber\\
&=\frac{\lambda^2}{2\hbar v_F^2}\int\limits_{-\infty}^\infty\frac{\mathrm{d}\epsilon}{2\pi\hbar}\int\limits_{-\infty}^\infty\frac{\mathrm{d}\epsilon^\prime}{2\pi\hbar}\frac{\cosh\left(\beta\epsilon\right)}{\cosh\left(\beta\epsilon\right)+\cosh\left(\beta\epsilon^\prime/2\right)}\frac{1}{i\hbar\omega_n-2\epsilon}\nonumber\\
&=\frac{\lambda^2}{\pi\hbar^2v_F^2}\int\limits_{-\infty}^\infty\frac{\mathrm{d}\epsilon}{2\pi\hbar}\frac{\epsilon}{\tanh\left(\beta\epsilon\right)\left(i\hbar\omega_n-2\epsilon\right)},
\end{align}
where we wrote $\epsilon_{k^\prime}\rightarrow-\epsilon_{k^\prime}$ in the second line. In order to deal with the remaining UV divergence, we reintroduce the energy cut-off $\Lambda$. Noting that the real part of the integrand is odd in $\epsilon$, we obtain
\begin{equation}
\Sigma(i\omega_n)=-\frac{i\omega_n\lambda^2}{2\pi^2\hbar^2v_F^2}\int\limits_{-\Lambda}^\Lambda\mathrm{d}\epsilon\,\frac{\epsilon}{\tanh\left(\beta\epsilon\right)}\frac{1}{(\hbar\omega_n)^2+(2\epsilon)^2},\label{eq:selfenergy}
\end{equation}
which diverges logarithmically as $\Lambda\rightarrow\infty$.

We now return to the autocorrelator from Eq.~(\ref{eq:corrD}), replacing $D_{bb}(i\omega_n)$ with $D_{bb}^\text{full}(i\omega_n)$ and evaluating the momentum integral. Using the same methods for dealing with momentum integrals as before, we find
\begin{align}
K^\tau_{cc}(i\Omega_{n>0},T)&=\frac{ie^2\Gamma}{4\hbar^2\beta}\sum\limits_{n^\prime=0}^\infty\left(D_{bb}^\text{full}(-i\omega_{n^\prime-n})-D_{bb}^\text{full}(i\omega_{n^\prime+n})\right)\nonumber\\
&=K^\tau_{cc}(i\Omega_{n>0},T)\Big|_{\lambda=0}-\frac{ie^2\Gamma}{4\hbar^2\beta}\sum\limits_{n^\prime=0}^\infty\Big(\big(D_{bb}(i\omega_{n^\prime-n})\big)^2\Sigma(i\omega_{n^\prime-n})+\big(D_{bb}(i\omega_{n^\prime+n})\big)^2\Sigma(i\omega_{n^\prime+n})\Big)+\mathcal{O}(\lambda^4),
\end{align}
where we used the fact that both $D_{bb}(i\omega_n)$ and $\Sigma(i\omega_n)$ are odd functions of $\omega_n$. Plugging in $D_{bb}(i\omega_n)$ and splitting the sum into several smaller sums, the lowest order correction to the current autocorrelator can be written as
\begin{align}
\Delta K^\tau_{cc}(i\Omega_{n>0},T)&=\frac{ie^2\Gamma}{2\beta}\left(\sum\limits_{n^\prime=0}^\infty\frac{\Sigma(i\omega_{n^\prime})}{\left(\hbar\omega_{n^\prime}+\Gamma\right)^2}-\sum\limits_{n^\prime=0}^{n-1}\frac{\Sigma(i\omega_{n^\prime})}{\left(\hbar\omega_{n^\prime}+\Gamma\right)^2}\right)\nonumber\\
&=\text{const.}-\frac{e^2\Gamma\lambda^2}{4\pi^2\hbar^3v_F^2\beta}\int\limits_{-\Lambda}^\Lambda\mathrm{d}\epsilon\sum\limits_{n^\prime=0}^{n-1}\frac{\epsilon}{\tanh(\beta\epsilon)}\frac{1}{\left(\hbar\omega_{n^\prime}+\Gamma\right)^2}\frac{\hbar\omega_{n^\prime}}{(\hbar\omega_{n^\prime})^2+(2\epsilon)^2}.
\end{align}
The sum over the Matsubara frequencies can be evaluated by using the partial fraction decomposition
\begin{align}
\frac{1}{\left(\hbar\omega_{n^\prime}+\Gamma\right)^2}\frac{\hbar\omega_{n^\prime}}{(\hbar\omega_{n^\prime})^2+(2\epsilon)^2}&=-\frac{1}{\left(\hbar\omega_{n^\prime}+\Gamma\right)^2}\frac{\Gamma}{\Gamma^2+(2\epsilon)^2}-\frac{1}{\hbar\omega_{n^\prime}+\Gamma}\frac{\Gamma^2-(2\epsilon)^2}{\big(\Gamma^2+(2\epsilon)^2\big)^2}\nonumber\\
&\quad\,+\frac{1}{\hbar\omega_{n^\prime}-2i\epsilon}\frac{1}{2\left(\Gamma+2i\epsilon\right)^2}+\frac{1}{\hbar\omega_{n^\prime}+2i\epsilon}\frac{1}{2\left(\Gamma-2i\epsilon\right)^2}
\end{align}
and applying the usual digamma function identities. Subsequently expanding the result to linear order in $\Omega_n$ and to lowest order in $1/\beta\Gamma$, we find
\begin{align}
\Delta K^\tau_{cc}(i\Omega_{n>0},T)&=\text{const.}-\frac{e^2\beta\Gamma\lambda^2}{32\pi^4\hbar^3v_F^2}\int\limits_{-\beta\Lambda}^{\beta\Lambda}\mathrm{d}(\beta\epsilon)\frac{\beta\epsilon}{\tanh(\beta\epsilon)}\nonumber\\
&\quad\,\times\left[\left(\psi^{(1)}\left(\frac{1}{2}-\frac{i\beta\epsilon}{\pi}\right)+\psi^{(1)}\left(\frac{1}{2}+\frac{i\beta\epsilon}{\pi}\right)\right)\frac{\hbar\Omega_n}{(\beta\Gamma)^2}+\mathcal{O}\big(\Omega_n^2,(1/\beta\Gamma)^3\big)\right].
\end{align}
Finally evaluating the remaining integral in the wide-band limit $\Lambda\rightarrow\infty$ and performing analytic continuation to real frequencies, we recover the lowest order correction to the dc linear susceptibility:
\begin{equation}
\Delta\chi_{cc}=-\frac{\pi^3e^2\lambda^2}{16h^3v_F^2}\frac{1}{\beta\Gamma}+\mathcal{O}\big((1/\beta\Gamma)^2\big).
\end{equation}
Identifying $1/\beta\Gamma$ as $T/T_K$, the charge conductance is therefore equal to
\begin{equation}
G=\frac{e^2}{2h}\left(1-\frac{\pi^3\lambda^2}{8h^2v_F^2}\frac{T}{T_K}+\ldots\right)\label{eq:Gpert}
\end{equation}
when approaching the local moment fixed point from below, where the dots contain all higher order terms in products of $\lambda^2$ and $T/T_K$; the leading order term $e^2/2h$ is the one from Eq.~(\ref{eq:cond}), and follows from evaluating Eq.~(\ref{eq:corrD}). This result for the leading order and the leading correction of the conductance $G$ agrees with the results from previous research~\cite{zheng2014pert,pustilnik2017pert,landau2018pert}. Note that the full linear term in $T/T_K$ is by itself a series in $\lambda^2$, while $\lambda$ is not necessarily small. Moreover, we see that the lowest order correction to the conductance vanishes as $T/T_K$ goes to zero, independent of $\lambda$. This is a manifestation of the irrelevance of the anisotropy $\Delta J_z\equiv J_z-J_\bot$: no matter the starting point (which is dictated by the parameter $\lambda$), the RG flow ensures that $\Delta J_z$ effectively goes to zero with the energy scale (in this case $T/T_K$), such that the EK point results become exact regardless of $\lambda$. 

\section*{Corrections to the Emery-Kivelson point heat conductance and its implications for the Wiedemann-Franz law}
One may treat finite-temperature corrections to the heat conductance in the vicinity of the EK point due to the leading RG irrelevant perturbations in a similar fashion. For heat conductance this involves evaluating five-loop diagrams -- a formidable task which we did not undertake in this work. However, it is clear from the structure of the perturbation theory that any correction in $\lambda$ is inevitably accompanied by powers of $T/T_K$. These are, after all, \emph{irrelevant corrections} to the fixed point properties, and do not affect the fixed point value itself as $T\rightarrow 0$ (this is the meaning of an irrelevant perturbation!). The form of the heat conductance, taking into account such corrections, would be of the generic form,
\begin{equation}
\kappa=\frac{\pi^2 k_{\text{B}}^2 T}{6h}\left(1-b \lambda^2 \frac{T}{T_K}+\ldots\right) \;,\label{eq:kappapert} 
\end{equation}
which is similar in structure to Eq.~(\ref{eq:Gpert}) for the charge conductance $G$.

In particular, note that we have a finite contribution to $\kappa$ when precisely \emph{at} the fixed point (obtained already at three-loop, as above). Similarly, the charge transport is finite at the EK fixed point. Therefore, when computing the WF ratio at the C2CK critical fixed point, we need only consider the values of $G$ and $\kappa$ at the EK point itself. Of course, this would be different if either $G$ or $\kappa$ were zero at the EK point, since then the leading corrections around the EK point would come into play. Fortuitously, this is not the case for the C2CK model at the critical point. 

In the main text of the paper, we discuss how at the Fermi liquid fixed points of the C2CK model, the charge and heat conductances vanish. In such situations, the WF ratio may still remain finite, however, due to the leading corrections to the fixed point values as the limit $T\rightarrow 0$ is taken. The WF law is in general violated in these circumstances. This is reminiscent of the calculation of the Wilson ratio at the EK point, which involves the ratio of the magnetic spin susceptibility and the heat capacity, which both vanish at the EK point. The proper calculation of the Wilson ratio therefore necessitates obtaining corrections to the EK point~\cite{sengupta1994ek}.

We again emphasize that this is \emph{not} necessary for the C2CK critical fixed point, because the EK values of $G$ and $\kappa$ are both finite.

\end{fmffile}

\end{document}